\documentclass[]{aa}
\usepackage{graphicx}
\usepackage{multirow}
 \usepackage{natbib,twoopt}
 
 \bibpunct{(}{)}{;}{a}{}{,}             
 
\usepackage{txfonts}
\usepackage[hyperindex,pdftex]{hyperref}
\newcommand{\chandra}{\it Chandra\rm}

\def\pks{PKS~2155-304}

\begin{document}

   \title{Discovery of Galactic \ion{O}{IV} and \ion{O}{V} X-ray absorption due to transition temperature gas in the \pks\ spectrum}

   \author{J. Nevalainen\inst{1}\fnmsep\thanks{jukka@to.ee}
          \and
           B. Wakker\inst{2}
          \and
         J. Kaastra\inst{3,4}
          \and
          M. Bonamente\inst{5}
          \and
           S. Snowden\inst{6}
          \and
         F. Paerels\inst{7}
          \and
           C. de Vries\inst{3}
}

\institute{Tartu Observatory, Observatooriumi 1, 61602 T{\~{o}}ravere, Estonia
\and
Supported by NASA/NSF, affiliated with Department of Astronomy, University of Wisconsin-Madison, Madison, WI 53706, USA
\and
SRON, Sorbonnelaan 2,  3584 CA Utrecht, Netherlands
\and
Leiden Observatory, Leiden University, PO Box 9513, 2300 RA Leiden, the Netherlands
\and
University of Alabama in Huntsville, Huntsville, AL 35899, USA
\and
NASA/Goddard Space Flight Center, Greenbelt, MD  20771, USA
\and
Columbia University, 1022 Pupin, 550 West 120th Street, New York, NY 10027, USA
}

   \date{Received ; accepted}

\abstract{
FUV observations have revealed the transition temperature gas (TTG; $\log{T({\mathrm K})}$ $\sim$5), located in the lower Galactic halo and in the High Velocity Clouds. However, the corresponding X-ray absorption has so far remained mostly undetected. In order to make an improvement in this respect in Galactic X-ray absorption studies, we accumulated very deep ($\sim$~3~Ms) spectra of the blazar \pks\ obtained with the spectrometers RGS1, RGS2, LETG/HRC and LETG/ACIS-S and studied the absorption lines due to the intervening Galactic components. The very high quality of the data and the coverage of important wavelengths with at least two independent instruments allowed us to reliably detect ten Galactic lines with better than 99.95\% confidence.

We discovered significant absorption from blended \ion{O}{IV} transitions 1s-2p $^2$S (22.571~\AA ), 1s-2p $^2$P (22.741~\AA ) and  1s-2p $^2$D (22.777~\AA ), and from the
\ion{O}{V} transition 1s-2p (22.370~\AA )
from TTG at $\log{T({\mathrm{K}})} \thinspace = \thinspace 5.2\pm0.1$. 
A joint X-ray and FUV analysis indicated that photoionisation is negligible for this component and that the gas is in a cooling transition phase.
However, the temperature is high enough that the column density ratio $N$(\ion{O}{IV})/$N$(\ion{O}{V}) is not significantly different from that in collisional ionisation equilibrium (CIE). Under CIE we obtained $N_{\mathrm{OIV}}$~= 3.6$\pm$2.0~$\times~10^{15}$~cm$^{-2}$, corresponding to $N_{\mathrm{H}}$~= 1.0$\pm$0.5~$\times~10^{19}~\frac{Z_{\odot}}{Z_{\mathrm{TTG}}}$~cm$^{-2}$. 
}


\titlerunning{Galactic \ion{O}{IV} and \ion{O}{V} discovered}

\authorrunning{J. Nevalainen et al.}

   \maketitle

\section{Introduction}
\label{intro}
The Galactic transition temperature gas (TTG; $\log{T({\mathrm{K}})}$ $\sim$ 5) has been very robustly detected via FUV absorption observations of ions like \ion{O}{VI}, \ion{N}{V}, \ion{C}{IV} and \ion{Si}{IV} \citep[e.g.][]{2012ApJ...749..157W}. Its kinematic structure is often complex due to the presence of the low and high velocity components.

The low velocity FUV-absorbing TTG component is likely located in the lower Galactic halo (see \citet{2012ApJ...749..157W} for the observations)
and can be understood via the Galactic Fountain model 
\citep[e.g.][]{1976ApJ...205..762S}. In this model the supernovae in the Galactic disk heat the interstellar medium to a temperature 
$\log{T(\mathrm{K})}$ $\sim$ 6, which consequently expands a few kpc away from the Galactic plane \citep[e.g.][]{2010PASJ...62..723H,2014PASJ...66...83S}, thus forming the hot phase of the Galactic halo. The cooling time scales are shorter than those required to reach a hydrostatic equilibrium and thus the cooled material falls towards the disk. The outcome of this process is a flow of material and regions containing TTG located in the lower Galactic halo.

Transition temperature gas at high velocities is likely more distant.
Some of these are low-metallicity accreting clouds, others are parts of the Magellanic stream, while still others are suspected of being outside the Milky Way's halo. TTG gas is expected to form around the cooler, denser, clouds seen in 21-cm emission as they interact with the hot coronal gas.

In X-rays, the observational view on the Milky Way is less complete.
The hot phase ($\log{T({\rm K})} \thinspace \sim$ \thinspace 6) of the Galactic halo has been detected via soft X-ray emission 
\citep[e.g.][]{1991Sci...252.1529S, 1991Natur.351..629B}
and via \ion{O}{VII} and \ion{O}{VIII} absorption \citep[e.g.][]{2015ApJS..217...21F}.
However, due to the relatively low efficiency and resolution of the high resolution X-ray spectrometers to date, the X-ray features of the FUV-detected transition temperature gas have not yet been previously detected.

Our aim in this work is to improve our insight into Galactic hot gas physics using soft X-ray spectroscopy, analysing very deep co-added exposures of the bright blazar PKS~2155-304. It is a calibration target for the XMM-Newton instruments RGS1 and RGS2 (collectively called RGS in the following) and for the Chandra LETG/HRC-S and LETG/ACIS-S  combinations (collectively called LETG in the following). Thus, this sight line through the Milky Way is covered by several independent high resolution X-ray data sets with very high statistical quality. We utilise these data for studying in detail the Galactic components in this direction via the X-ray absorption lines the Galactic medium imprints on the  PKS~2155-304 spectrum. Combining the X-ray and FUV data (see below), we assess the relative importance of collisional ionisation and photo-ionisation and to examine the thermal instability of TTG.

\section{Data}
The detection of most of the Galactic X-ray absorption lines is at the limit of the capabilities of the currently most powerful high-resolution X-ray spectrometers: XMM-Newton/RGS and Chandra/LETG.  The exposure time for even the brightest blazars like \pks\ should be around a Ms for robust Galactic spectroscopy using absorption lines at the typical equivalent width (EW) level of a few m\AA. Since Ms exposures are observationally too expensive for a single program, one can only achieve the required sensitivity level by combining a large number of individual observations of the same source. Since \pks\ is one of the brightest compact X-ray sources with a simple power-law emission spectrum, it has been frequently monitored by XMM-Newton/RGS and Chandra/LETG for calibration purposes. Thus the current \pks\ observation data base provides an opportunity to advance Galactic X-ray spectroscopy.

The co-addition of a large number of observations, particularly if they span a long period of time, is not trivial. The possible difference in the locations of the bad pixels at different times may create false line-like features if simply co-adding the individual spectra \citep[see][]{2011A&A...534A..37K}. This problem could be minimised, in principle, by jointly analysing the individual spectra without co-adding. This approach is problematic as well, due to the complexity of dealing with a large number of involved spectra. Also, the low signal-to-noise in the individual spectra complicates the estimation of the possible wavelength scale offset. Consequently the modelling of the lines may not be accurate.

\subsection{XMM-Newton/RGS}
\label{RGSdata}
In order to overcome these co-addition problems, Kaastra et al. 2015 ("Effective area calibration of the RGS",  SRON internal report) introduced a new method for combining a large number of RGS observations. They used all available data for \pks\  (see Table~\ref{rgs.tab}) for the purpose of RGS calibration. In the current paper we use this data set. The essential steps of the procedure are as follows. 

\begin{itemize}

\item
The spectra are extracted using the standard SAS tools; we use flux-calibrated spectra.

\item
Dead channels due to bad pixels were determined for each individual spectrum and their effect was removed before co-adding the spectra.

\item
The time-dependent corrections of the RGS efficiency are implemented by creating an effective area file for each individual spectrum according to its observation date.

\item

Each spectrum has been fitted with a broken power-law model, using a total
column density for neutral hydrogen of $N_{\mathrm{H}}$~=~1.24~$\times~10^{20}$~cm$^{-2}$ \citep{2011ApJ...728..159W}.
To model the effects of dust on the Fe-L and O-K edges we use the "amol"
model in SPEX, which uses typical (average) total metal abundances for the
amount of O, Mg, Si, and Fe in the gas phase of 0.849, 0.187, 0.162 and 0.130
times solar \citep{2013A&A...551A..25P, 2012A&A...539A..32C}. We use the properties of MgSiO$_3$ and solid iron to modify the
O-K and Fe-L absorption edges, since these molecules seem to give the best
match to the data \citep{2012A&A...539A..32C}.

\item 
The final spectrum is constructed by co-adding the weighted residuals at each spectral bin using the individual spectral fits above.
Stacking the residuals is equivalent to adding observations, assuming that the continuum modelling is accurate.

\end{itemize}

\begin{figure}
\vspace{-6.5cm}
\includegraphics[width=11cm,angle=0]{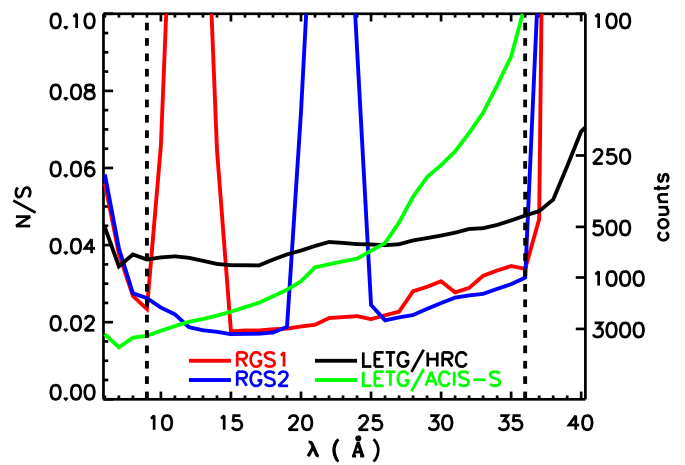}
\caption{The ratio of statistical uncertainties to the signal per bin in the co-added PKS~2155-304 observations using RGS1 1st order (red), RGS2 1st order (blue), LETG/HRC-S (black) and LETG/ACIS-S (green). The RGS bin size is 20~m\AA\ while LETG  bin size is 25~m\AA. The curves have been smoothed strongly for clarity. The dashed lines indicate the waveband selected for the analysis in this paper.}
\label{sens.fig}
\end{figure}

The above procedure has been carried out separately for the RGS1 and RGS2 data, keeping the 1st and 2nd orders separate. This resulted in four spectra, 
each with a total exposure time of $\sim$1.2~Ms, corrected for the absorption and the continuum. The spectra have a constant spectral bin size of 20~m\AA, i.e. the energy resolution has been oversampled by a factor of $\sim$~3. Due to the high exposure time of the spectra the number of counts in 20~m\AA\ channels exceeds a few 1000 and thus the statistical uncertainties are at the $\sim$~2--3\% level of the source emission in such bins (see Fig.\ref{sens.fig}). In the current paper we started from the RGS1 and RGS2 spectra, reduced as described above.

Given the large number of counts in our RGS spectral bins (this is also true for LETG, see below) we used $\chi^2$ statistics to determine the best-fitting  parameters for the baseline absorption line model and the uncertainty in the parameters. The average RGS relative effective area uncertainty from
one resolution element to another in our adopted waveband is $\sim$2\% (J. Kaastra,  arXiv:1611.05924) and thus the systematic uncertainties may play a significant role in our analysis. Since the RGS resolution element size is 60~m\AA , a 2\% feature in the continuum corresponds to an equivalent width of $\sim$1~m\AA ,
which we take as a rough estimate of the calibration uncertainty effect on our EW measurements. 
We will check the LETG results with a similar level of systematic uncertainty.
A more advanced error propagation method is desirable. However, we do not know the probability distributions of the systematic uncertainties in our data. Thus, in the basic analysis we compare results obtained with only statistical uncertainties with those adding the above uncertainty in quadrature to the statistical ones, and discuss the effects.

We subtracted the background produced by the SAS package.  The background flux is $\sim$1\% of the source emission, i.e. at the level of the statistical uncertainties of the total signal. Thus, if the background is uncertain by a given fraction, the background-subtracted signal is uncertain by the same fraction of the statistical uncertainties, i.e. negligible.
Thus, we subtracted the standard background (also in the case of LETG) and did not propagate any background-subtraction related uncertainties to our results. 

\subsection{Chandra/LETG}
We accumulated all the publicly available \chandra\ LETG data on \pks\  (see Table~\ref{letg.tab}). The total useful exposure times are $\sim$870~ks and $\sim$310~ks for the  ACIS-S/LETG and HRC-S/LETG combinations.

Since the data reduction procedure used for our XMM-Newton data is not yet applicable to the Chandra data, we reduced the Chandra data using the standard publicly available tools  in CIAO 4.7.  The data were processed with the standard pipeline (\textit{chandra\_repro}) that generates source spectra, background spectra and response files, separately for the +1 and -1 order data. We co-added the individual spectra, keeping the ACIS and HRC data separate. 
In order to reach a sensitivity similar to that of the RGS, we further combined the +1 and -1 order spectra, using the \textit{combine\_grating\_spectra} tool. The analysis of \chandra\ data is therefore performed on two spectra, one containing the LETG/ACIS data and the other one containing the LETG/HRC data.
We binned the spectra using a constant spectral bin size of 25~m\AA , similar to that used for RGS.

\subsection{Sensitivities of different instruments}
\label{sens}
When comparing the results obtained with different instruments, we need to consider the variation of their sensitivity and wavelength coverage. The sensitivity of a given telescope/spectrometer/detector combination to detect absorption lines with a given data set depends primarily on the number of observed counts,
i.e. on the product of the effective area, the exposure time and the average emission level of PKS~2155-304. All of these factors related to the different data sets used in this work vary and hence does the sensitivity. We estimated the relative sensitivity as a function of wavelength by comparing the ratios of the statistical uncertainties to the signal per adopted bin size (see Fig.~\ref{sens.fig}). 

In most of the common wavebands covered with different instruments, the 1st order data of RGS1 and RGS2 are the most sensitive. In a statistical sense, they are capable of detecting features whose depth exceeds 2-3\% of the continuum level in the 9-36~\AA\ band we selected for the analysis in this work. 
The sensitivity of the LETG/HRC-S given purely by the number of counts is smaller than that of the RGS. However, the better energy resolution of the LETG (FWHM~$\sim$~40~m\AA) compared to the RGS  (FWHM~$\sim$~60~m\AA) improves the relative sensitivity of the LETG/HRC-S towards that of the RGS. The LETG/ACIS-S has a similar sensitivity as RGS at $\lambda$~=~12~\AA , and it 
decreases significantly with the wavelength. Since the 2nd order RGS data are of much lower sensitivity, and their wavelength coverage is limited, we do not include them in the basic analysis. However, we will utilise the second order data for testing the robustness of our results considering the instrument calibration uncertainties (see Section~\ref{2ndorder}).

\section{Method for finding Galactic lines}
\label{velocity}
In this work we adopted a very conservative approach for the line identification. 
By utilising the very high signal-to-noise of our data and the coverage of the important wavelengths by more than one instrument, 
we attempted to produce a very reliable set of Galactic line measurements in this work. 

We tested the presence of the strongest a priori known Galactic absorption lines by examining the wavebands containing the ground state lines from \cite{1996ApJ...465..487V} and the strongest inner transitions included in the SPEX distribution \citep{1996uxsa.conf..411K}. We analysed the 9--36~\AA\ band data of each instrument, i.e. RGS1, RGS2, LETG/HRC-S and LETG/ACIS-S using the X-ray spectral analysis package SPEX~3.00.00 \citep{1996uxsa.conf..411K}. However, we did not consider features at the wavelengths of the known bad channels caused by CCD gaps in RGS data or strong instrumental problems at $\lambda\,\sim\,23\,$\AA\ in both the RGS and the LETG (see the XMM-Newton Users Handbook\footnote{https://heasarc.gsfc.nasa.gov/docs/xmm/uhb/}  and  The Chandra Proposer's Observatory Guide\footnote{http://cxc.harvard.edu/proposer/POG/html/}). The only exception is the case for \ion{O}{II} which we treat separately in section~\ref{OII}.

We performed the analysis in a narrow band of $\sim$~1~\AA\ centered at a given a priori line.
We modelled the LETG local blazar emission continuum using a power-law model, allowing both the photon index and the normalisation to be free parameters. In the case of the RGS, the continuum has already been modelled and its effect has been removed from the data (see Section~\ref{RGSdata}). Thus, when analysing the lines in the RGS data, we applied a constant model for the continuum, allowing the level to vary from unity, in order to allow for the statistical uncertainties of the continuum.

We fitted the line data with a ``slab'' model of SPEX \citep{2002A&A...386..427K} which calculates the transmission of a thin slab of material whose ion column densities can be varied independently, i.e. the ion ratios are not determined by the ionisation balance. Since we are studying the strongest Galactic lines, with $\tau_0$ of order 1, saturation effects are important and we properly take them into account with our procedure (i.e. we do not assume linear growth of N(ion) as a function of EW).

The model produces the Lorentz profile for each transition in the SPEX atomic data base for a given ion, including the Auger broadening.\footnote{The damping constant $a$ for the strongest lines (\ion{O}{VII} and \ion{O}{VIII} Ly$\alpha$) is below 0.01, i.e. the damping-wing effect is negligible.}
Thus, the calculations are also accurate in the case of blended multiplets, like \ion{O}{IV}. 
The Gaussian component of the Voigt profile is calculated based on the input value of the total velocity dispersion (thermal and non-thermal). \cite{2007ApJ...665..247W} preferred a Doppler parameter of $\sim$~50~km~s$^{-1}$ when jointly fitting the LETG/ACIS-S and LETG/HRC-S data of the OVII lines in the \pks\ data. 
While our data are of higher statistical quality,  our attempt to constrain the velocity components did not yield useful results.  
Thus, when using the Voigt profile of the ``slab'' model, we allowed the total velocity dispersion to vary in a range 20-35~km~s$^{-1}$, 
which includes the non-thermal broadening observed in FUV (Wakker et al., in prep.) and thermal broadening of 10~km~s$^{-1}$ (the value for oxygen at 
$\log{T(K)}$ = 5). 
If the tested line-like feature was acceptably fitted by a known transition of the expected ion at z = 0, we upgraded it as a Galactic line candidate.

We then used the best-fit model to obtain the column density of the given ion and the EW of the corresponding line (or blend). We noticed that using instead a single Gaussian profile for the 
\ion{O}{I}, \ion{O}{IV}, \ion{O}{VIII} and \ion{C}{VI} multiplets yielded a significantly lower EW value, compared to the one obtained with ``slab''. 
This is due to the very high statistical quality of our data rendering the Gaussian approximation of the blended multiplets inaccurate. 
We obtained the statistical uncertainties of the column densities by $\chi^2$ minimisation and consequently the constraints EW$\pm \sigma_{EW,stat}$ for the equivalent width.
We added the approximate calibration uncertainty of 1~m\AA\ (see section~\ref{RGSdata}) in quadrature to the above uncertainties to obtain the total uncertainties
$\sigma_{EW,tot}$. We used the ratio EW / $\sigma_{EW,tot}$ as a measure of the detection significance N$_\sigma$ (see Table~\ref{gal_lines.tab}).

\begin{table*}
 \centering
    \caption{The Galactic absorption line measurements}
 \label{gal_lines.tab}
  \begin{tabular}{llc|cc|cc|cc|cc}
  \hline\hline
                             &             &        &              &      &             &     &              &     &              &      \\
Ion       & \multicolumn{2}{c}{Transition} & \multicolumn{2}{c}{RGS1 1st}   & \multicolumn{2}{c}{RGS2 1st}   & \multicolumn{2}{c}{LETG/HRC-S}   &  \multicolumn{2}{c}{LETG/ACIS-S}     \\
          & name & $\lambda_{0}$\tablefootmark{a} (\AA) & EW (m\AA )\tablefootmark{b}  & N$_\sigma$\tablefootmark{c}  & EW (m\AA )\tablefootmark{b}  & N$_\sigma$\tablefootmark{c} & EW (m\AA )\tablefootmark{b}  & N$_\sigma$\tablefootmark{c} & EW (m\AA )\tablefootmark{b}  & N$_\sigma$\tablefootmark{c} \\
                               &             &        &              &      &             &     &              &     &              &           \\ 
\ion{Ne}{IX}                   & 1s-2p       & 13.447 & n/c          & --   & $5.1\pm1.5$ & 3.4 & 8.0$\pm$2.1  & 3.9 & $\le$2.0     & n/d \smallskip \\
\ion{O}{VII}                   & 1s-3b       & 18.628 & 4.7$\pm$1.5  & 3.2  & $3.0\pm1.5$\tablefootmark{d} & 2.0\tablefootmark{d}   & 2.6$\pm$1.9 & 1.3 & 4.6$\pm$1.5  & 3.0 \smallskip \\
\multirow{2}{*}{\ion{O}{VIII}} & \multirow{2}{*}{1s-2p} & 18.967 & \multirow{2}{*}{9.0$\pm$1.7} &  \multirow{2}{*}{5.2} & \multirow{2}{*}{ 5.5$\pm$1.6} & \multirow{2}{*}{5.7} & \multirow{2}{*}{7.8$\pm$2.4} & \multirow{2}{*}{3.3} & \multirow{2}{*}{5.3$\pm$1.7} & \multirow{2}{*}{3.2} \\
                              & & 18.972 &              &      &             &     &              &     &              &          \smallskip \\ 
\ion{O}{VII}                 & 1s-2p       & 21.602 & 15.4$\pm$1.5 & 10.1 & n/c         & --  & 11.3$\pm$2.6 & 4.3 & 10.0$\pm$1.7 & 6.0     \smallskip  \\
\ion{O}{V}                   & 1s-2p       & 22.370 & 3.0$\pm$1.5  & 2.0  & n/c         & --  & 3.7$\pm$2.3  & 1.6 & $\le$1.1  & n/d \smallskip \\
\multirow{3}{*}{\ion{O}{IV}} & 1s-2p $^2$S & 22.571 & \multirow{3}{*}{7.0$\pm$2.8} &  \multirow{3}{*}{2.5} & \multirow{3}{*}{n/c} &  \multirow{3}{*}{--} & \multirow{3}{*}{8.2$\pm$3.1} & \multirow{3}{*}{2.6} & \multirow{3}{*}{$\le$4.2} &  \multirow{3}{*}{n/d} \\
                             & 1s-2p $^2$P & 22.741 &              &      &             &     &              &     &              &           \\ 
                             & 1s-2p $^2$D & 22.777 &              &      &             &     &              &     &              &         \smallskip  \\
\multirow{2}{*}{\ion{O}{I}}  & \multirow{2}{*}{1s-2p} & 23.510 & --          &  --  &  --         & -- & \multirow{2}{*}{17.3$\pm$3.3} & \multirow{2}{*}{5.2} &  \multirow{2}{*}{9.0$\pm$2.0} & \multirow{2}{*}{4.5} \\
                             &                        & 23.511 &              &      &             &     &              &     &              &           \\ 
\ion{N}{VI}                  & 1s-2p       & 28.788 & 4.5$\pm$2.0\tablefootmark{d}  &  2.3\tablefootmark{d} & 3.1$\pm$1.6 & 1.9 & 5.0$\pm$2.4  & 2.1 & $\le$3.2     & n/d \smallskip \\  
\ion{N}{I}                   & 1s-2p       & 31.286 & --           &  --  & --          & --  & 10.6$\pm$2.9  & 3.6 &  $\le$6.1     & n/d  \smallskip \\ 
\multirow{2}{*}{\ion{C}{VI}} & \multirow{2}{*}{1s-2p} &  33.734 & \multirow{2}{*}{ 7.7$\pm$2.6\tablefootmark{d}} & \multirow{2}{*}{ 2.9\tablefootmark{d}} & \multirow{2}{*}{8.8$\pm$2.0} & \multirow{2}{*}{4.3} & \multirow{2}{*}{3.6$\pm$2.6} & \multirow{2}{*}{1.4} & \multirow{2}{*}{$\le$6.1} & \multirow{2}{*}{n/d} \\  
   &               & 33.740      &              &      &             &     &              &     &              &           \\ 
                           &              &        &              &      &             &     &              &     &              &           \\ 
\hline\hline                  
\end{tabular}
\tablefoot{ \\
\tablefoottext{a}{A priori values from \cite{1996ApJ...465..487V}, \cite{1996uxsa.conf..411K} and \cite{2005ApJ...627.1066G}.} \\
\tablefoottext{b}{Equivalent width and total 1~$\sigma$ uncertainties (i.e. statistical ones and systematic 1~m\AA )  for detected lines. For the non-detections (n/d) the upper 1~$\sigma$ limit is given. n/c means that the line is not covered by the instrument.}\\
\tablefoottext{c}{Significance of the line detection in terms of 1$\sigma$ uncertainty.}\\
\tablefoottext{d}{The problematic single channel has been omitted when deriving the values.}
}
\end{table*}

\subsection{Problems and solutions}
The identification of the Galactic lines based on their wavelength is complicated due to the uncertainties in the determination of the wavelength of a given line-like feature in the data. The statistical uncertainties around the peak of the line may cause a random apparent shift of the line centroid. Systematic line shifts may be introduced by the uncertainties due to the limitation of the accuracy of the wavelength scale calibration, i.e. 6~m\AA\ for the RGS and 10~m\AA\ for the LETG.
To evaluate the effect of the apparent centroid shifts we also fitted the data of our Galactic line candidates using Gaussians with free centroids. 
A comparison between the fitted and a priori centroid wavelengths of our Galactic line candidates indicated no systematic bias in the wavelength scales. 
{\it We accepted such lines whose fitted centroid wavelengths were consistent with one of the tested a priori lines, considering the uncertainties of the centroid wavelength at 95\% CL.}

In a few cases there were larger shifts in the wavelength, amounting up to 60~m\AA\ for \ion{O}{VIII} and \ion{N}{VI} (RGS1) and \ion{O}{VII}~Ly$\beta$ (RGS2).
This may be related to the observed non-statistical fluctuations in RGS data \citep{2007ApJ...656..129R}. 
Since these well-known Galactic lines were securely identified with other instruments, we additionally accepted these into our list of Galactic line candidates. We will pay special attention to the effect of these shifts of our reported results below.

In addition to the redshift issue, we have several other problems when identifying and interpreting the line-like like features. Since the expected absorbers have line widths much smaller than the resolution element, there may be incidences where the stacking procedure leads to incorrect results. Also, given the high statistical precision of a few per cent of the blazar emission level in our adopted spectral bins with sizes of 20-25~m\AA, we must be very careful not to interpret small residuals due to the possible effective area calibration uncertainties above this level as astrophysical signals. Furthermore, since some of the predicted Galactic lines are relatively weak, we must minimise the possibility of interpreting statistical fluctuations at a few per cent level as celestial. In order to minimise the above effects, we performed screening of the line-like features by applying the following two criteria.

\begin{enumerate}

\item
{\it Statistical significance level of 95\%.}\\
We estimated the statistical significance of a given line using the best-fit value and the statistical uncertainties of the equivalent width of the line, keeping the centroid wavelength fixed to an a priori value.
We examined a total band width of $\sim$30 \AA . With a resolution of  $\sim$ 50 m\AA, we have  $\sim$600 resolution elements.
Thus, we expect to have  $\sim$30 random events exceeding the CL = 95\% level (2$\sigma$)  in the full spectrum.
Since we examine features in bands of width  $\sim$ 1 \AA, the misidentification is possible if one of the 30 2$\sigma$  fluctuations hits the 1 \AA\  search region.
Since our search region is $\sim$3\% of the full band, the expected number of random fluctuations per band is below 1. 
Thus, we rejected such lines whose line flux deviates from zero at less than 95\% confidence level i.e. 2$\sigma$. The only exception is the 1.6$\sigma$ HRC detection of OV, which we additionally accepted (see below).

\item
{\it Detection with relevant instruments.}\\
Even if a given Galactic line candidate was detected significantly and consistently with the atomic physics, there is still some probability of a misidentification of a statistical fluctuation or a calibration related feature as a Galactic line. In order to minimise this probability, we utilised the four independent instruments. If a given astrophysical line is covered with several instruments of comparable sensitivity, it should be detected with all at a comparable confidence level. 
Given the problems with ACIS detections (see below) we applied this requirement only to RGS and HRC.
Given that RGS1 and RGS2 are the most sensitive instruments in most of the studied waveband, we first required that the given line-like feature in our improved list of Galactic line candidates (see above) must be detected by both RGS1 and RGS2 at 95\% CL, if covered with both. In case the given feature is covered with only one of the RGS units, we conservatively\footnote{Since the LETG sensitivity is lower than that of RGS, we may reject some of the true Galactic lines with this strict requirement} required that LETG/HRC-S detects the same feature at 95\% CL.

The probability of the null hypothesis, i.e. the no-line assumption for the detection of a feature with two independent instruments, is (1-0.95)$^2$ = 0.0025. Thus, the null hypothesis can be discarded at the 1-0.0025=99.75\% level if two independent lines are detected at the 2$\sigma$ CL.
If the feature also passed this test, we upgraded it as a true Galactic line.  

\end{enumerate}

\begin{figure*}
\vspace{-13cm}
\includegraphics[width=18cm,angle=0]{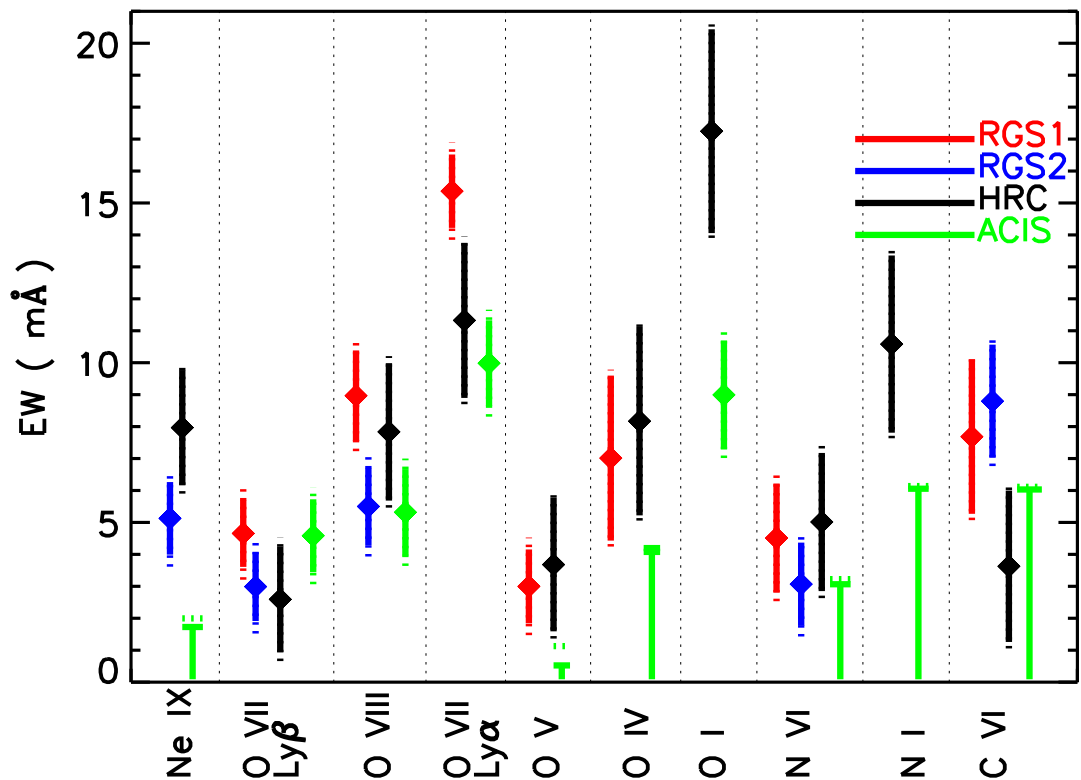}
\caption{The equivalent widths and the statistical 1~$\sigma$ uncertainties of lines (solid lines) detected with RGS1 (red), RGS2 (blue), LETG/HRC-S (black) and LETG/ACIS (green). For the non-detections, we report the 1~$\sigma$ upper limits. The dotted lines indicate the effect of adding the 2\% systematic uncertainties.
\label{ew.fig}}
\end{figure*}

\begin{figure*}
\vspace{-12cm}
\includegraphics[width=18cm,angle=0]{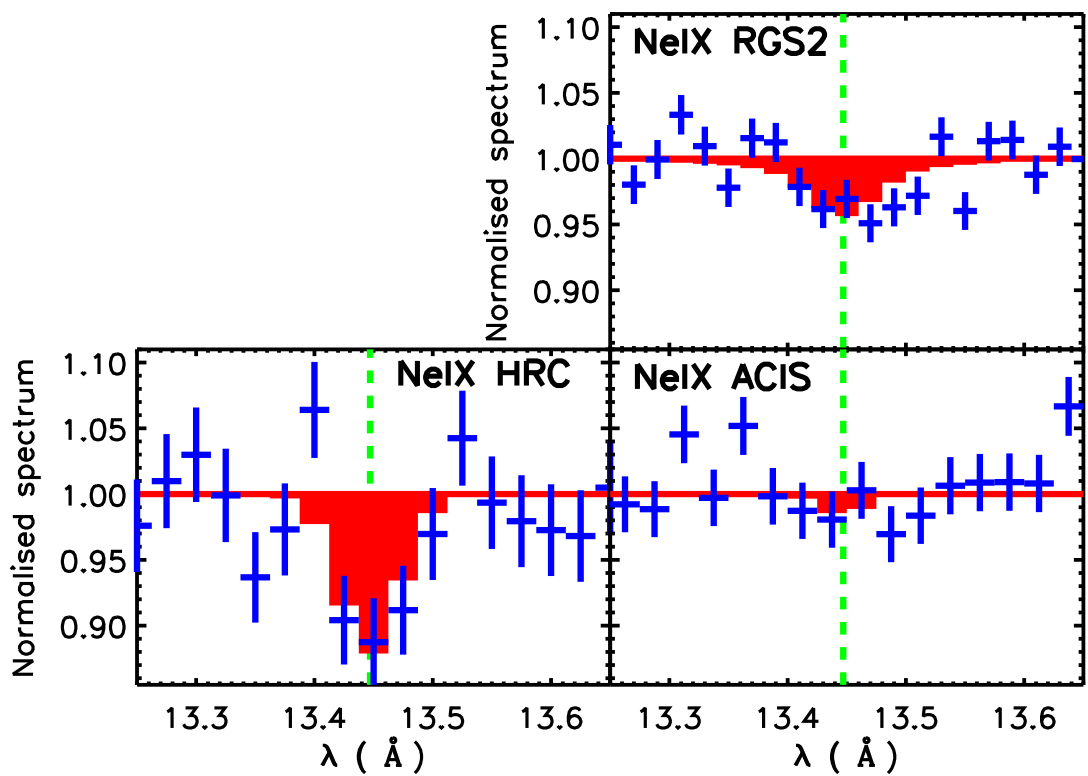}
\caption{The normalised data (blue crosses) and the best-fit models (solid red lines) of the \ion{Ne}{IX} line for RGS2 (upper right), LETG/HRC-S (lower left) and LETG/ACIS-S (lower right). The RGS1 does not cover these wavelengths.
The normalisation is done by dividing the spectra by the power-law component. The error bars reflect only the statistical uncertainties.
}
\label{NeIX.fig}
\end{figure*}

\begin{figure*}
\vspace{-12cm}
\includegraphics[width=18cm,angle=0]{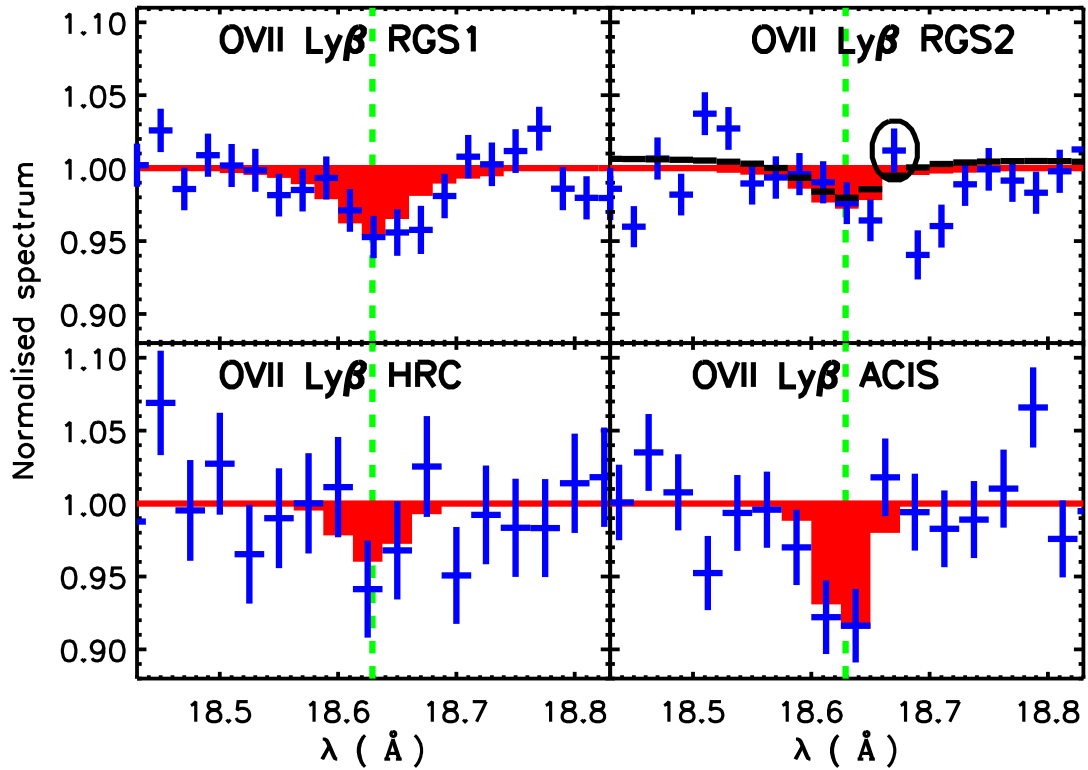}
\caption{As Fig.~\ref{NeIX.fig}, but for \ion{O}{VII} Ly$\beta$ transition line. The black line in the top right panel indicates the best-fit line profile when excluding the strongly deviant channel (surrounded by a black ellipse).
}
\label{OVII_beta.fig}
\end{figure*}

\begin{figure*}
\vspace{-12cm}
\includegraphics[width=18cm,angle=0]{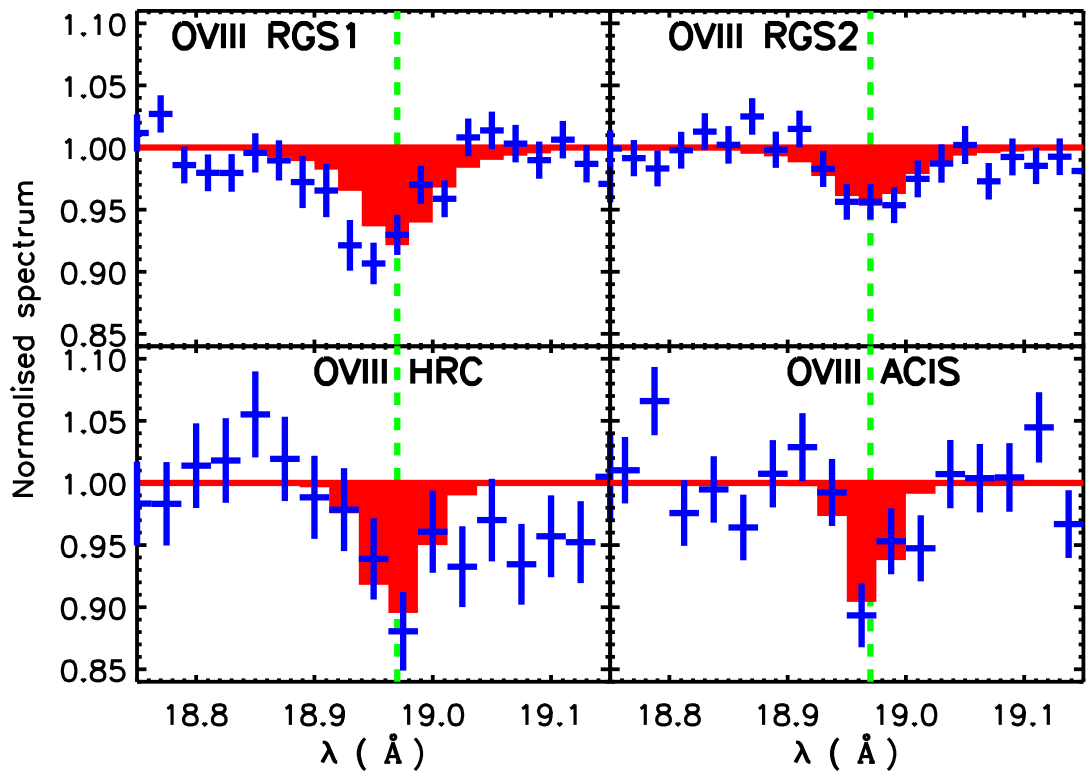}
\caption{As Fig.~\ref{NeIX.fig}, but for the \ion{O}{VIII} Ly$\alpha$ transition line.
}
\label{OVIII.fig}
\end{figure*}

\begin{figure*}
\vspace{-11cm}
\includegraphics[width=18cm,angle=0]{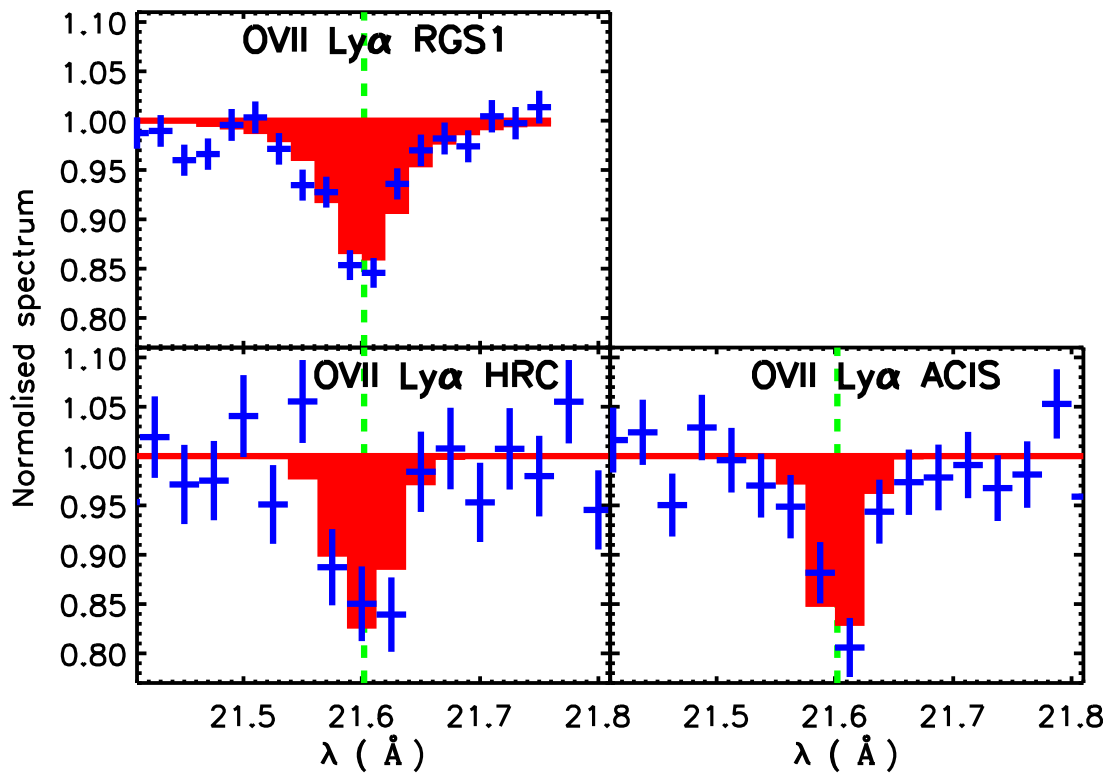}
\caption{As Fig.~\ref{NeIX.fig}, but for the \ion{O}{VII}~Ly$\alpha$ line. The RGS2 does not cover these wavelengths. 
}
\label{OVII_alpha.fig}
\vspace{8cm}
\end{figure*}

\begin{figure*}
\vspace{-12cm}
\includegraphics[width=18cm,angle=0]{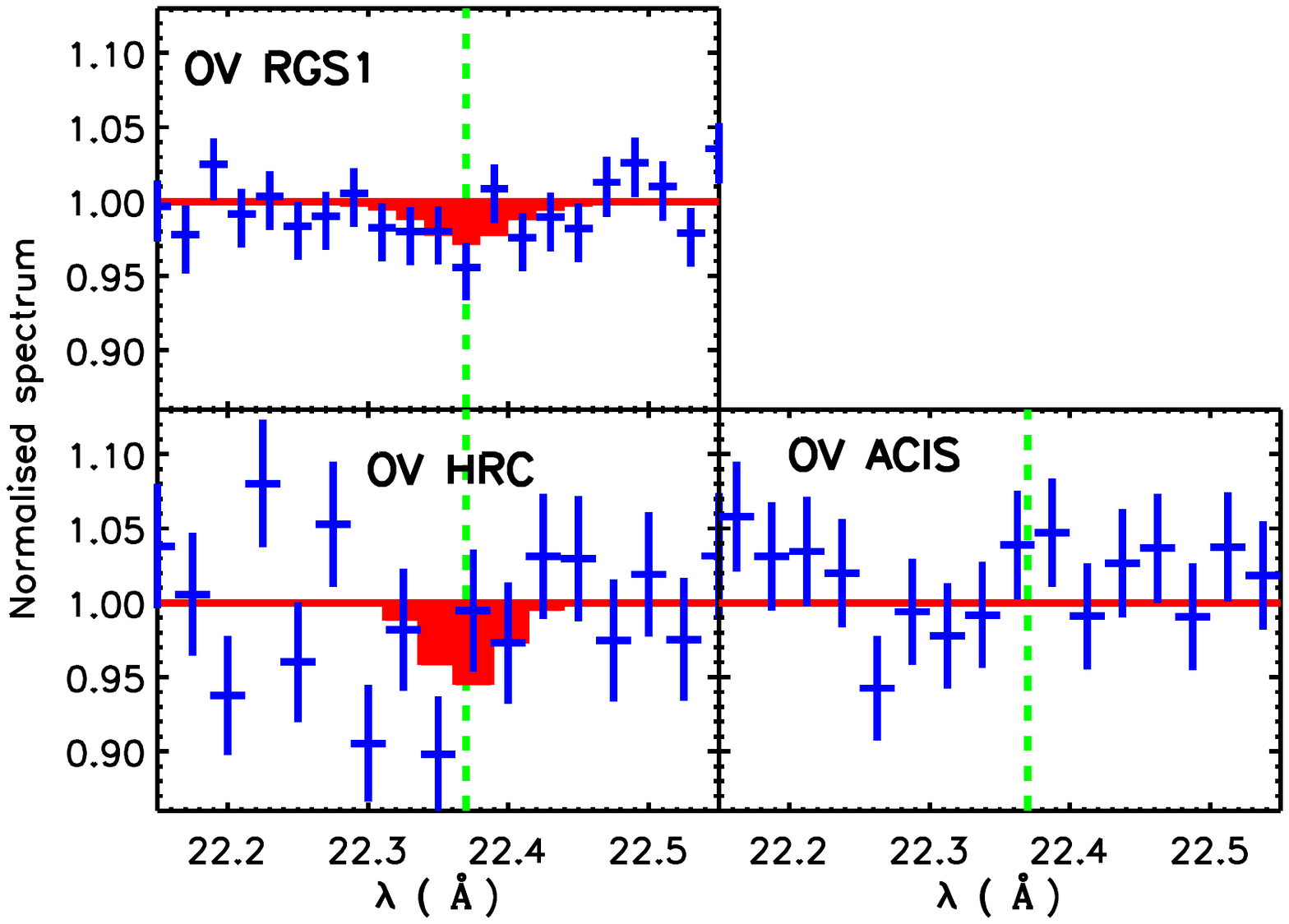}
\caption{
As Fig.~\ref{NeIX.fig}, but for the \ion{O}{V} line. RGS2 does not cover these wavelengths.}
\label{OV.fig}
\end{figure*}

\begin{figure*}
\vspace{-12cm}
\includegraphics[width=18cm,angle=0]{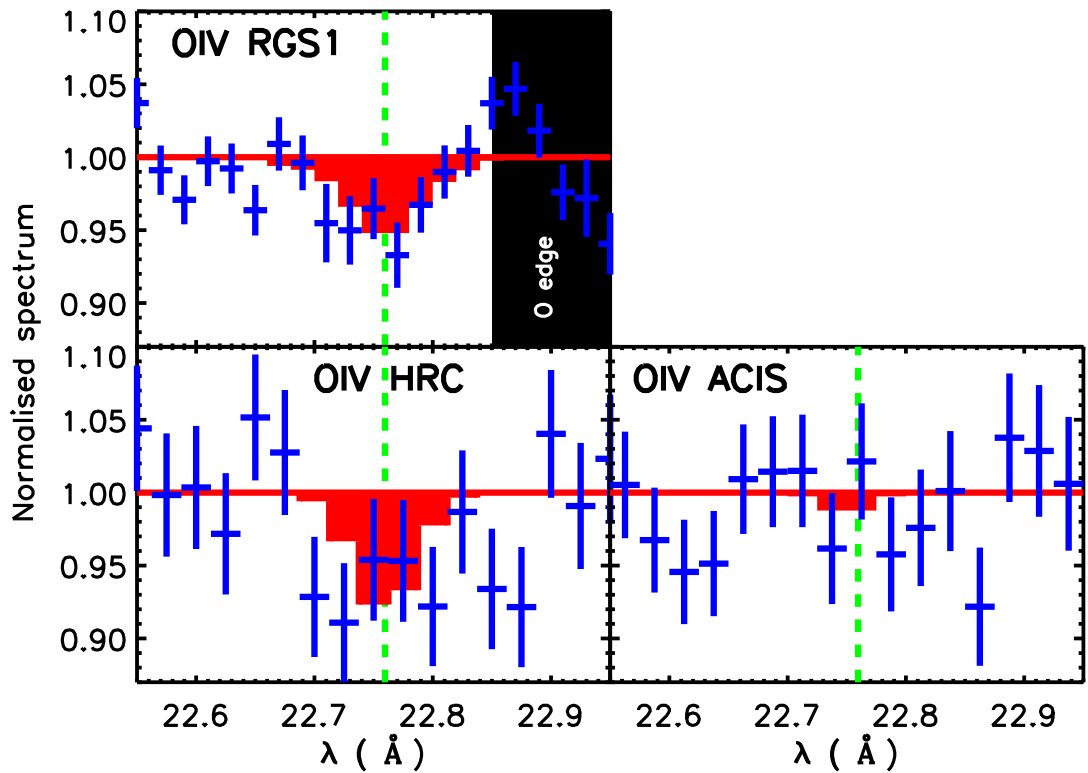}
\caption{
As Fig.~\ref{NeIX.fig}, but for the \ion{O}{IV} line. RGS2 does not cover these wavelengths. 
The RGS1 oxygen edge is indicated with the black box.}
\label{OIV.fig}
\end{figure*}

\begin{figure*}
\vspace{-12cm}
\includegraphics[width=18cm,angle=0]{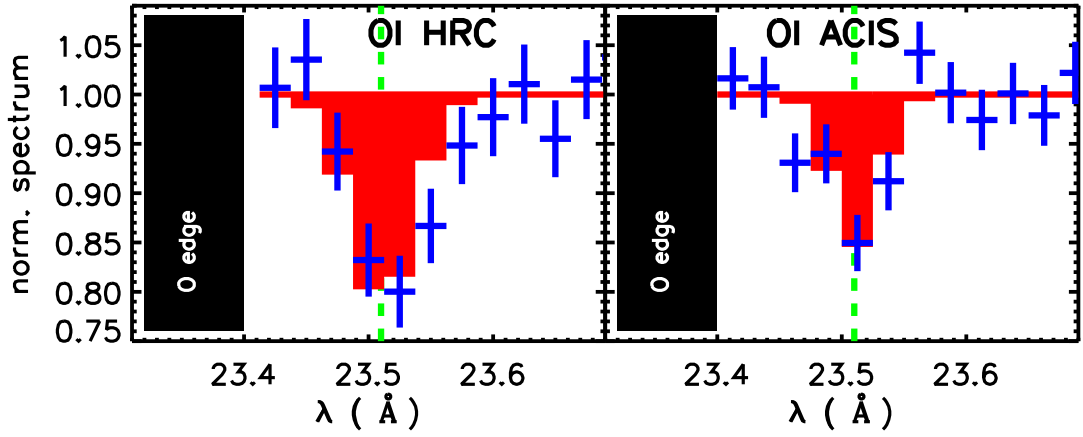}
\vspace{-4cm}
\caption{As Fig.~\ref{NeIX.fig}, but for the \ion{O}{I} line. The black boxes indicate the omitted wavelengths due to the instrumental oxygen edge. 
}
\label{OI.fig}
\end{figure*}

\begin{figure*}
\vspace{-12cm}
\includegraphics[width=18cm,angle=0]{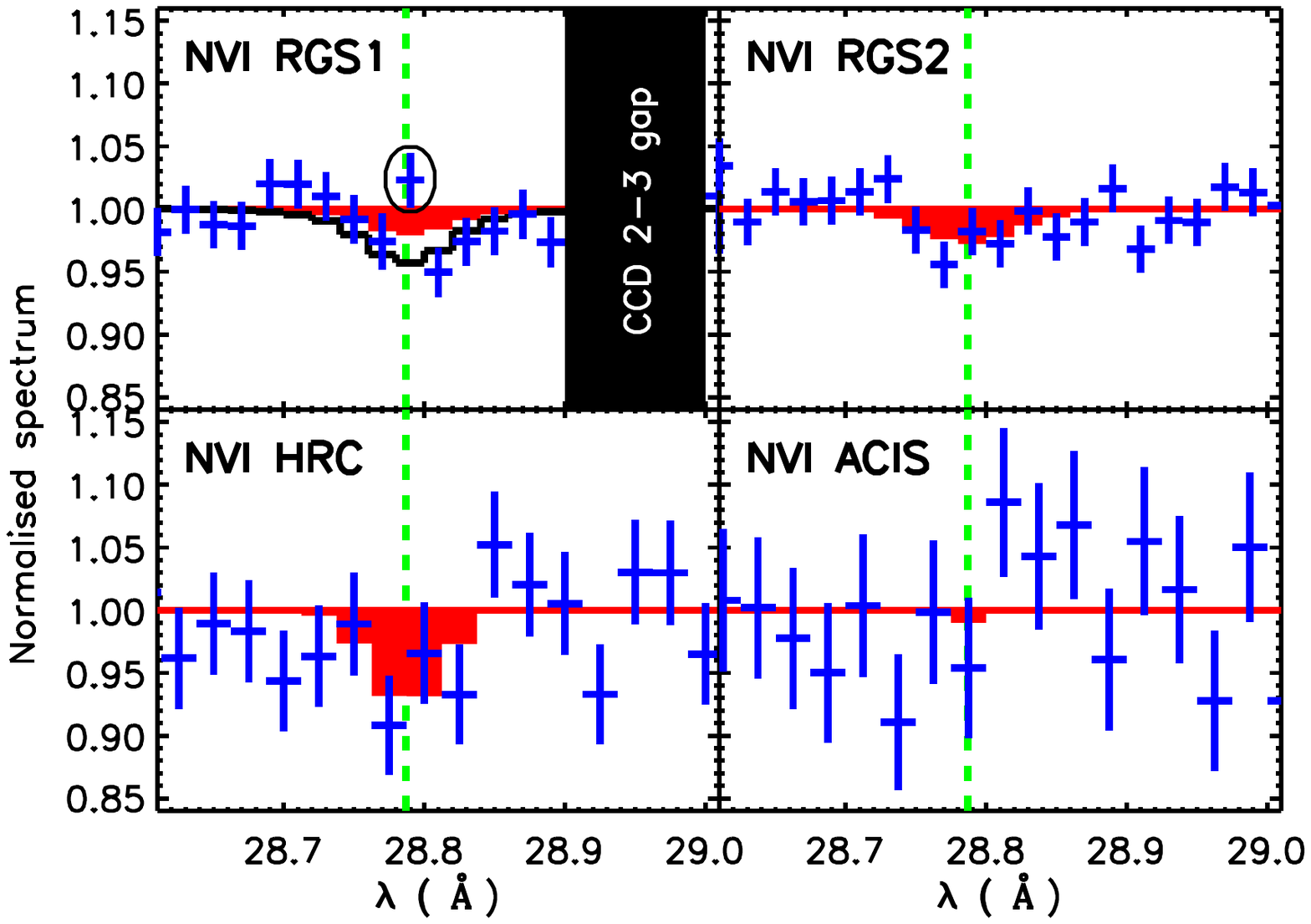}
\caption{As Fig.~\ref{NeIX.fig}, but for the \ion{N}{VI} line. The gap between the RGS1 CCD 2 and CCD 3 gap is denoted with the black box. The black line in the top left panel indicates the best-fit line model when excluding the strongly deviant channel (surrounded by a black ellipse).}
\label{NVI.fig}
\vspace{4cm}
\end{figure*}

\begin{figure*}
\vspace{-12cm}
\includegraphics[width=18cm,angle=0]{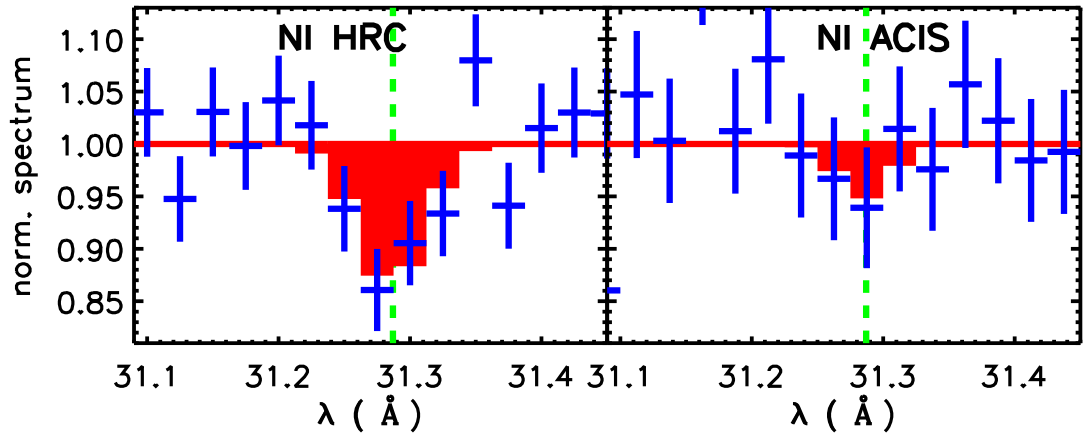}
\vspace{-4cm}
\caption{As Fig.~\ref{NeIX.fig}, but for \ion{N}{I} line.  
}
\label{NI.fig}
\end{figure*}

\begin{figure*}
\vspace{-12cm}
\includegraphics[width=18cm,angle=0]{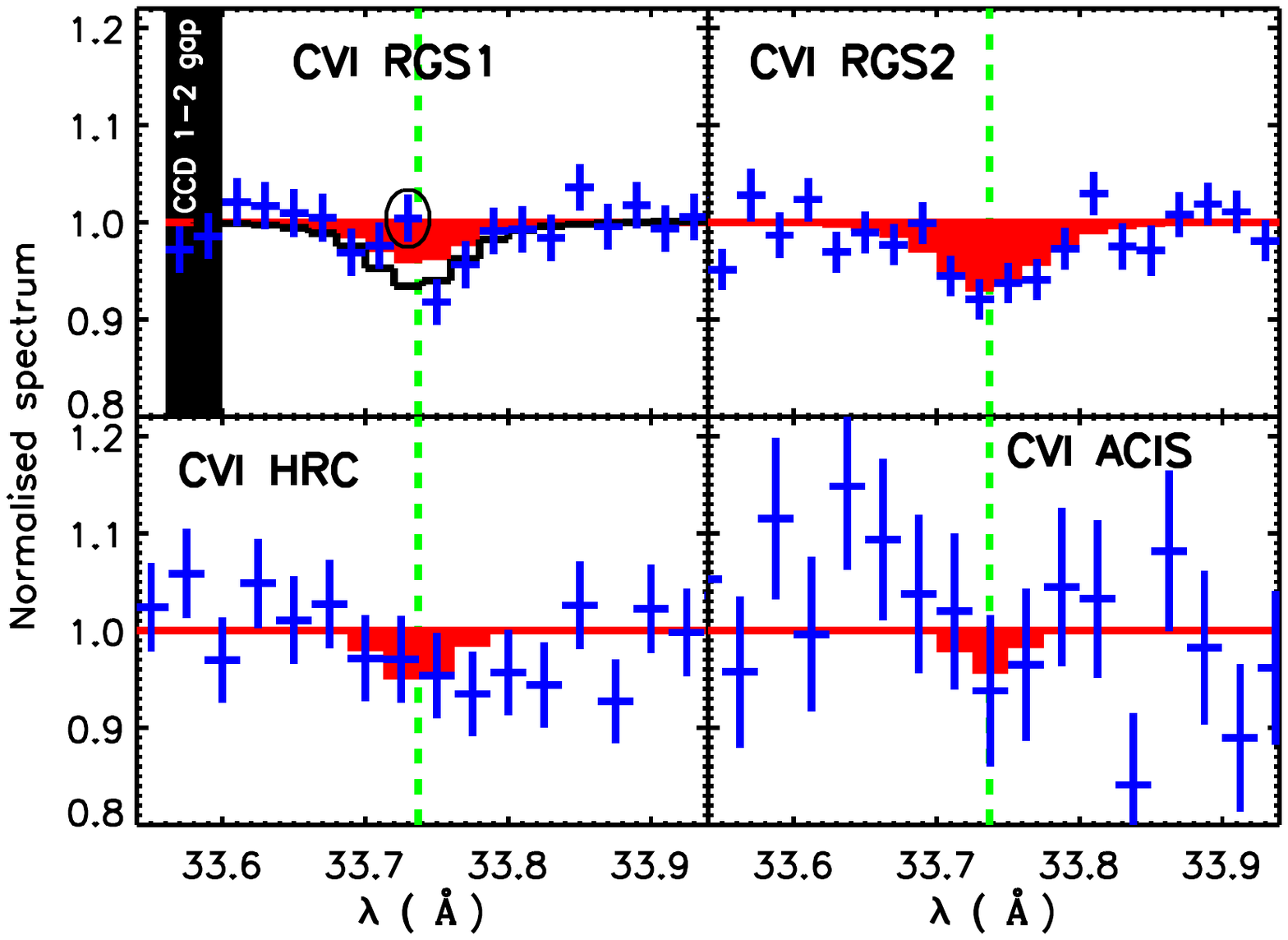}
\caption{As Fig.~\ref{NeIX.fig}, but for the \ion{C}{VI} line. The gap between the RGS1 CCD 1 and CCD 2 is denoted with a black box. The black line in the top left panel indicates the best-fit line model when excluding the strongly deviant channel (surrounded by a black ellipse).  
}
\label{CVI.fig}
\vspace{4cm}
\end{figure*}

\section{Detections and identifications}
\label{components.sec}

The above procedure resulted in detections of ten Galactic absorption lines identified with their ionic species and transitions (see Table~\ref{gal_lines.tab} and Figs.~\ref{ew.fig}-\ref{CVI.fig}). As expected by the sensitivity calculations (see Fig.~\ref{sens.fig}), RGS and HRC detected lines with equivalent width exceeding $\approx$~2~m\AA . Considering all the instruments, there were 25 independent significant detections.
The RGS1 and RGS2 1$\sigma$ EW constraints for all four lines in common agreed.  
In most cases the RGS and HRC measurements yielded EW values consistent for a given line within the uncertainties. 
The only exception is \ion{N}{VI}, whose measurements by RGS2 and HRC differ marginally. 
The systematic uncertainty level of 1~m\AA\ causes a very small effect to the uncertainties (see Fig.~\ref{ew.fig}).
In summary, the possible systematic uncertainties of the effective area calibration of RGS and HRC do not yield significant biases for the EW measurement.

However, the ACIS seems to differ systematically from the other instruments.
We discuss this issue next, together with other sources of systematic uncertainties. 

\subsection{Instrumental issues}
\label{instr}

\subsubsection{Centroid shifts}
When using the best-fit centroid wavelengths to identify the Galactic lines (Section~\ref{velocity}), we noted significant shifts for 
the centroids of the \ion{O}{VIII} and \ion{N}{VI} (RGS1) and \ion{O}{VII}~Ly$\beta$ (RGS2) lines from the a priori values.
It is not clear whether the true EW is recovered better by using the apparent or the a priori wavelength for the line centroid. 
Thus, we examined the EW values obtained by keeping the centroid free or fixing to a priori value.
The change was not significant, implying that whatever is causing the shift of these lines, does not have significant effect in our EW 
measurements.

\subsubsection{Non-statistical fluctuations}
\label{RGS2probs}
In case of RGS1 lines \ion{N}{VI} and \ion{C}{VI} and RGS2 \ion{O}{VII}~Ly$\beta$ line, the best-fit model 
yielded significant residuals (see Figs.~\ref{NVI.fig}, \ref{CVI.fig} and \ref{OVII_beta.fig}).
In very deep exposures such as the ones in this work, occasionally non-statistical fluctuations become visible (see Rasmussen et al., 2007). Usually, these are confined to excursions in a single narrow wavelength bin (significantly narrower than the spectrometer resolution). A single such excursion seems to be visible in the data of all the above lines between wavelengths
28.782-28.795,  33.721-33.74 and 18.660-18.680~\AA . When ignoring these, the EW of RGS1  \ion{N}{VI} and \ion{C}{VI} lines increased significantly, into better agreement with RGS2. The RGS2 OVII~Ly$\beta$ was not affected. In the following, we exclude these problematic channels.

\subsubsection{LETG/ACIS-S negative bias}
\label{ACIS_instr}
At $\lambda~\le 25$~\AA\ the LETG/ACIS-S should be more sensitive than LETG/HRC-S  and comparable to RGS (see Fig.~\ref{sens.fig}). In fact, the detection confidences of the \ion{O}{VII}  and \ion{O}{VIII}  1s-2p lines for ACIS are greater than for HRC (see Table~\ref{gal_lines.tab}). 

However, ACIS did not detect the \ion{Ne}{IX}, \ion{O}{V}, \ion{O}{IV}, \ion{N}{VI}, \ion{N}{I} and  \ion{C}{VI} lines, which were detected with other instruments covering these wavelengths. 
Also, the ACIS detection of \ion{O}{I} yielded a value of $\sim$8~m\AA\ for EW(\ion{O}{I}), lower by a factor of 1.9 than that of the HRC (see Fig.~\ref{OI.fig} and Table~\ref{gal_lines.tab}). These measurements point to a negative bias of several m\AA\ in LETG/ACIS-S EW measurements.
This effect may be related to the pixelation of the ACIS CCDs slightly undersampling the Chandra PSF. While the native spectral bins used in the standard data processing are smaller than the energy resolution, the final effective resolution may be worse than that of HRC, washing out the weakest features. 

On the other hand, the LETG/ACIS did detect the \ion{O}{VII}~Ly$\alpha$, \ion{O}{VII}~Ly$\beta$ and \ion{O}{VIII}~Ly$\alpha$ lines and yielded EW consistent with the RGS2 and the HRC for these lines. Thus, the systematic effect is hard to understand and to take properly into account when measuring absorption lines with LETG/ACIS-S.

\subsection{Results}
\label{res.sec}
Our very strict criteria for the identification of the Galactic lines were needed to minimise the possibility of misidentification of noise as Galactic lines. The downside is that we may omit some interesting weaker lines due to the Milky Way (e.g. Nicastro et al., 2016) or the intervening extragalactic Warm Hot Intergalactic Medium \citep[e.g.][]{2007ApJ...670..992F,2007ApJ...665..247W}. We will address these in a future paper. 

Considering the temperatures where the absorption lines from different ions reach their maximal equivalent width \citep[e.g.][]{2008SSRv..134..155K}, our detections indicated three distinct absorbing components: Neutral disk (ND), hot halo (HH) and Transition Temperature Gas (TTG). 
\begin{itemize}

\item
{\bf ND:} The \ion{O}{I} and \ion{N}{I} ions (detected with the LETG) are obviously associated with the neutral $kT$\;$\sim$\;$10^{-3}$\;keV\;($\log{T({\mathrm{K}})}$\;$\sim$\;4) Galactic disk (ND) absorber. \\

\item
{\bf HH:} An absorber with $kT$\;$\sim$\;0.1\;keV\;($\log{T({\mathrm{K}})}$\;$\sim$\;6) is indicated by the lines from \ion{Ne}{IX}, \ion{O}{VII},  \ion{O}{VIII},  
\ion{N}{VI} and \ion{C}{VI}. We attribute this to the hot part of the Galactic Halo (HH). \\

\item
{\bf TTG:} The detection of the lines due to the inner shell \nolinebreak{1s-2p} transitions of \ion{O}{V} and \ion{O}{IV}
with RGS1 and LETG/HRC-S (see Figs.~\ref{OV.fig} and \ref{OIV.fig} ) revealed the existence of an absorber with $kT$\;$\sim$\;0.01\;keV\;($\log{T({\mathrm{K}})}$\;$\sim$\;5), i.e. Transition Temperature Gas (TTG). This is the first time that this component has been seen in X-ray absorption, and the first time that \ion{O}{IV} and \ion{O}{V} have been
definitely detected in the Milky Way. 

We note that a study of the RGS effective area calibration around the oxygen edge \citep{2003A&A...404..959D} reported an absorption feature at 22.77~\AA\ in the RGS1 spectra of PKS~2155-304 and Mrk~421. Since the same feature was indicated in the Chandra LETG data of the same sources, and it was absent in the RGS data of the Galactic sources Sco-X1 and 4U~0614+091, they concluded that the feature is a true Galactic signal in the directions towards PKS~2155-304 and Mrk421. They suggested that this feature is due to absorption by \ion{O}{IV}. 
Our analysis confirms this suggestion, and we also detect \ion{O}{V} from the same absorber. 

\end{itemize}

\begin{figure*}[t]
\vspace{-13cm}
\includegraphics[width=18cm,angle=0]{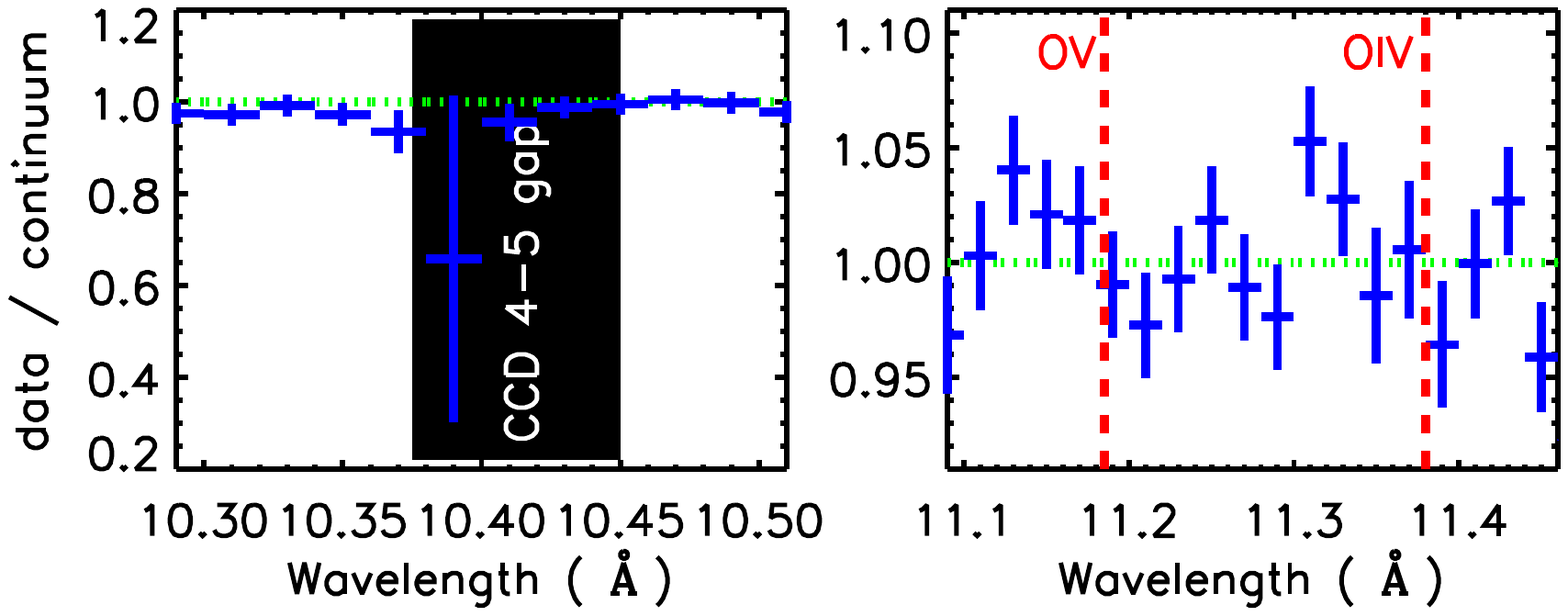}
\vspace{-4cm}
\caption{The RGS1 2nd order data of PKS~2155-304, divided by the continuum, are shown with blue crosses in different wavebands (different panels).
The black box is placed at 0.5 times the wavelengths of the CCD gaps in the RGS1 1st order (left panel). The vertical lines are placed at 0.5 times the wavelengths of \ion{O}{IV} and \ion{O}{V} (right panel).
}
\label{ccdgaps.fig}
\end{figure*}

\begin{figure*}
\vspace{-9cm}
\includegraphics[width=12cm,angle=0]{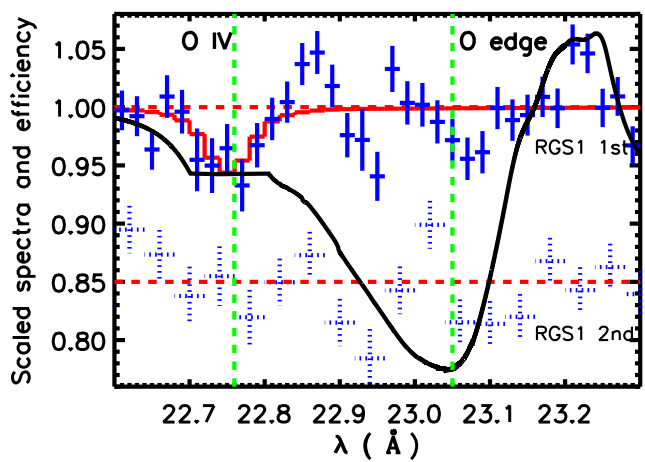}
\caption{
The RGS1 1st order data (blue solid crosses) and the best-fit power-law + narrow Gaussian fit (solid red line), divided by the power-law component in the waveband containing the \ion{O}{IV} line and the oxygen edge (green dashed lines)
The RGS1 2nd order data (blue dotted crosses) are displayed at wavelengths 2 times the original ones, scaled arbitrarily for display purposes. 
The black curve indicates the RGS1 instrumental efficiency from \cite{2003A&A...404..959D}, scaled to unity at $\lambda$~=~22.6~\AA\ .}
\label{OIV_Oedge.fig}
\end{figure*}

\subsubsection{Second order test for the TTG lines}
\label{2ndorder}
Given that our Galactic \ion{O}{V} and \ion{O}{IV} X-ray absorption detections are the first secure ones in the literature (except for the suggestion in \citet{2003A&A...404..959D}), and that we for the first time use these lines to detect the Galactic transition temperature gas in X-rays,  we utilised the second order RGS1 data for additional testing and confirmation. We used the fact that a detector position of a given wavelength of the first order corresponds exactly to half of that wavelength in the second order. Thus, a given instrumental artifact at a detector which gives a feature at the first order wavelength $\lambda_1$ should yield a feature in the second order spectrum at wavelength $\lambda_2~=~0.5~\times~\lambda_1$. In fact, the dead areas between the CCD gaps in the 1st order RGS1 spectra imprint a sudden drop of counts in narrow bands in the 2nd order data at half the wavelengths of the 1st order CCD gaps (see Fig.~\ref{ccdgaps.fig}).

We found that at the wavelengths 0.5 times of those of \ion{O}{IV} and \ion{O}{V}, the 2nd order data are consistent with a flat continuum, i.e. strong detector artifacts erroneously causing the \ion{O}{IV} or \ion{O}{V} detections are ruled out (see Fig.~\ref{ccdgaps.fig}).

\subsubsection{\ion{O}{IV} and the oxygen edge}
As stated before, we have not considered features at wavebands significantly affected by known instrumental effects (CCD gaps and O edge). 
Given that the wavelength of the \ion{O}{IV} line ($\lambda$\thinspace$\approx$\thinspace22.76\thinspace\AA\ ) is close to the band affected by the RGS1 instrumental oxygen edge, we made an additional check. The instrumental efficiency of RGS1 varies rapidly by $\sim$20\% in narrow wavebands at $\lambda$\thinspace=\thinspace22.8--23.3\thinspace\AA\  (see Fig.~\ref{OIV_Oedge.fig}). In this band the residuals indicate calibration-related features at the 5\% level of the continuum. The 2nd order data also indicate features at these wavelengths, demonstrating that uncertainties of the efficiency calibation around the oxygen edge may amount to deviations at the level of 5\% of the continuum. At the wavelengths of the \ion{O}{IV} line (22.7 - 22.8~\AA ) the RGS1 efficiency is flat and thus likely more accurately calibrated. In fact, the data are well fitted with the power-law + narrow Gaussian model, with no significant scatter (see Figs.~\ref{OIV.fig}~and~\ref{OIV_Oedge.fig}). Thus, the calibration uncertainties due to the oxygen edge around the \ion{O}{IV} line are likely smaller than 5\%, i.e. they cannot produce an artificial line-like feature similar to the one we associate with
\ion{O}{IV}.

\subsection{OII~1s-2p}
\label{OII}
Inspired by the recent report on the detection of Galactic \ion{O}{II}~1s-2p line \citep{2016MNRAS.457..676N} with LETG in the \pks\ sight line we checked this line in our spectra.
We had excluded these wavelengths ($\lambda$~$\sim$~23.35~\AA ) from our previous analysis due to the coincidence of the wavelengths of \ion{O}{II}~1s-2p with an instrumental feature in RGS1 and LETG. \cite{2003A&A...404..959D} argued that in both instruments there is \ion{O}{I}~1s-2p ($\lambda$~=~23.5~\AA ) absorption, shifted to 23.35~\AA\ due to \ion{O}{I} in the solid compounds of the instruments (metal oxide or water ice). This feature is included in the effective area calculations of both RGS and LETG.

\begin{figure}
\vspace{-17cm}
\includegraphics[width=24cm,angle=0]{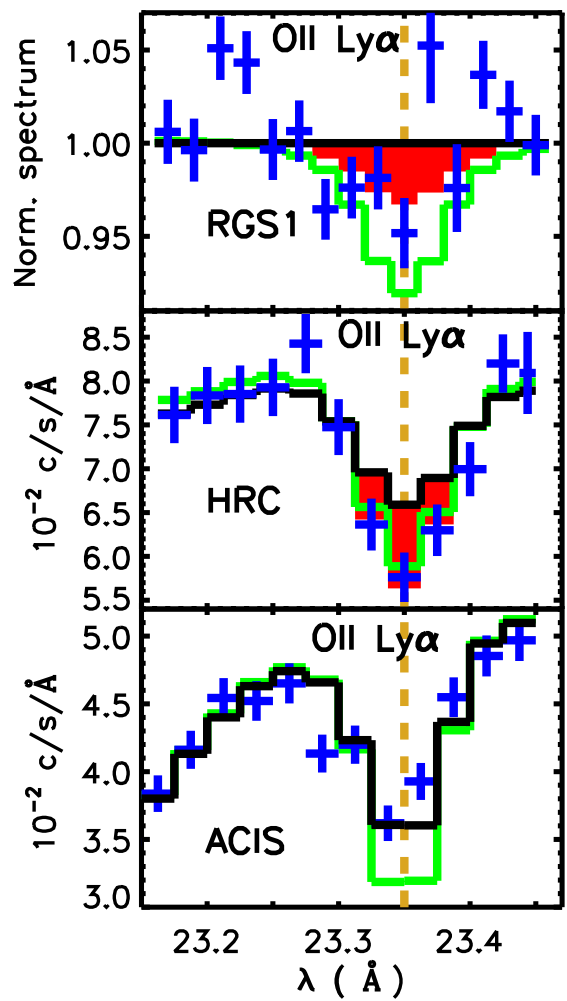}
\caption{The blue crosses show the data around the \ion{O}{II}~1s-2p wavelengths for RGS1 (upper panel), HRC (middle panel) and ACIS (lower panel). In case of HRC and ACIS we show the data without dividing with the continuum while this is not possible with RGS data due to the special data processing. The black lines indicate the continuum, in case of HRC and ACIS convolved with the responses. The red areas (for RGS1 and HRC) indicate the Gaussian model contribution above the continuum. In case of ACIS it is negligible. The green curves indicate the model from Nicastro et al. (2016).}
\label{OII_alpha.fig}
\end{figure}

We fitted a narrow Gaussian (FWHM = 10 km s$^{-1}$) plus a local power-law in the $\sim$~1~\AA\ band centered at $\lambda$~=~23.35~\AA , independently to RGS1, HRC and ACIS data.  We ignored the waveband 23.45--23.60~\AA\ in order not to bias the continuum due to the \ion{O}{I} line. In the case of the RGS1 and the HRC, the data indicated significant excess absorption on top of the instrumental feature (see Fig.~\ref{OII_alpha.fig}). The Gaussian modelling yielded inconsistent values EW(RGS1)~=~3.2$\pm$1.3~m\AA\ and EW(HRC)~=~11.6$\pm$2.5~m\AA\ for the excess. ACIS yielded only an upper limit EW(ACIS)~$\le$~0.5~m\AA , inconsistent with the other instruments. This is different from \cite{2016MNRAS.457..676N} who reported that HRC and ACIS yielded consistent EW for \ion{O}{II}.
The average value of EW(\ion{O}{II}~1s-2p) measured with HRC and ACIS in \cite{2016MNRAS.457..676N} is 8.6$\pm$1.1~m\AA\ which is inconsistent with our RGS1 and ACIS values. The origin of the discrepancy between our measurements and those of \cite{2016MNRAS.457..676N} is unclear.

A very generous maximum EW for \ion{O}{II} at $\lambda$ = 23.7~\AA\ can be estimated by assuming that all oxygen for the amount of $N$(\ion{H}{I}) in the \pks\ sight line 
($\sim10^{20}$~cm$^{-2}$) is in the form of \ion{O}{II}. Assuming additionally a turbulent broadening of 60 km~s$^{-1}$ yields EW(\ion{O}{II})\;$\le$\;9~m\AA .
However, in reality the radiation field conditions in the disk and inner halo indicate that the \ion{O}{II} to \ion{O}{I} column density ratio is much below 1.
Thus, the expected EW for \ion{O}{II} is below 1~m\AA , much smaller than that derived with RGS1 and HRC.

\begin{table}
 \centering
  \caption{\ion{O}{II} measurements
  \label{OII.tab}}
    \begin{tabular}{lc}
  \hline\hline\\
                     & EW(\ion{O}{II} 1s-2p)  \\
                     & m\AA\              \\
RGS1                 & 3.2$\pm$1.3        \\
HRC                  & 11.6$\pm$2.5       \\
ACIS-S               & $\le$~0.5          \\
N16\tablefootmark{a} & 8.6$\pm$1.1        \\
\hline\hline
\end{tabular}
\tablefoot{\\
\tablefoottext{a}{N16: average of HRC and ACIS-S as reported by \cite{2016MNRAS.457..676N}}} 
\end{table}

Our low ACIS value compared to that of RGS1 could be understood by the under-performance of ACIS we found above. However, the significantly high HRC EW measurement, compared to RGS1 is inconsistent with our work. Namely, six of the secure Galactic lines are detected with both RGS1 and HRC, within statistical precision of 2~m\AA , and there is no instance where HRC would yield significantly higher EW than RGS1. 

Thus, assuming that the calibration of the instrumental feature at $\lambda$~=~23.35~\AA\  is accurate, our measurements indicate that the reported OII~1s-2p signal \citep{2016MNRAS.457..676N}  is not due to a constant astrophysical source, as it should yield a constant EW for the same source as measured with different instruments. Also our HRC and RGS1 measurements of the EW 
are much bigger than allowed by the above  \ion{H}{I} consideration.
Thus, it is likely that the signal is at least partly due to inaccuracies of the calibration of the instrumental feature at 23.35~\AA . The simultaneous calibration of the instrumental feature and the measurement of the possible astrophysical \ion{O}{II} signal requires more work.

\section{Column densities without thermal equilibrium assumptions}
\label{NOCIE}

\begin{figure*}
\vspace{-12cm}
\includegraphics[width=18cm,angle=0]{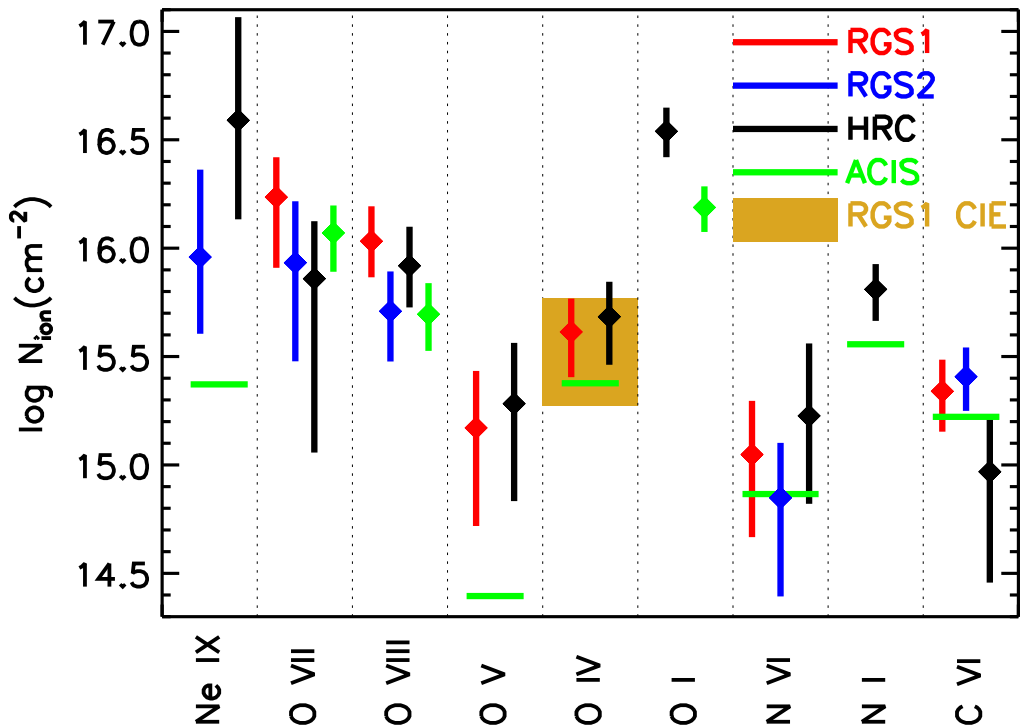}
\caption{Best-fit values and confidence intervals including statistical and systematic uncertainties at 1$\sigma$ level of the column densities of the ions producing the reliable detected Galactic lines (diamonds and vertical bars).  The upper limits obtained with ACIS are indicated with green horizontal bars. Values are derived without an assumption on the ionisation balance (solid lines) 
except for \ion{O}{IV}, for which we additionally show the RGS1 values assuming CIE (filled rectangle).}
\label{col.fig}
\end{figure*}

The line fitting procedure described above also yielded the column densities of the ions producing the lines
(see section~\ref{velocity}).
As indicated by the consistence of EW measurements above,
for each ion, RGS1, RGS2 and HRC yielded consistent column densities (see Table~\ref{EWtau.tab} and Fig.~\ref{col.fig}).

In the case of the TTG, the RGS1 constraints at the 1$\sigma$ (statistical and systematic) uncertainty level (EW$_{\mathrm{OIV}}$\;=\;7.0$\pm$2.8\;m\AA\ and 
EW$_{\mathrm{OV}}$\thinspace=\;3.0$\pm$1.5\;m\AA ) 
correspond to 
$\log{N(\mathrm{OIV}(\mathrm{cm}^{-2}))}$ = $15.61^{+0.15}_{-0.21}$
and
$\log{N(\mathrm{OV}(\mathrm{cm}^{-2}))}$ = $15.17^{+0.26}_{-0.45}$

The upper limits of the column densities derived using the ACIS data
are inconsistent with the constraints derived with other instruments (see Table~\ref{EWtau.tab} and Fig.~\ref{col.fig}) for several ions.
Due to the several non-detections of the lines, we did not use the ACIS data in the spectral analysis (section~\ref{ciemeth}).

\begin{table}
 \centering
  \caption{Ion column densities} 
\scriptsize{
  \label{EWtau.tab}
    \begin{tabular}{lcccc}
  \hline\hline
        &      &                &          &                             \\ 
Ion     & \multicolumn{4}{c}{$\log{N({\mathrm{cm}}^{-2})}$\tablefootmark{a}}                     \\
                              &  RGS1                 &  RGS2                  &  HRC         &  ACIS           \\ 
                              &                       &                        &              &                 \\ 
\ion{Ne}{IX}                  & --                    & 15.96$^{+0.40}_{-0.35}$   & 16.59$^{+0.48}_{-0.46}$   & $\le$~15.37     \smallskip \\ 
\ion{O}{VII\tablefootmark{b}} & 16.24$^{+0.18}_{-0.33}$  & 15.93$^{+0.28}_{-0.45}$   & 15.86$^{+0.27}_{-0.80}$  & 16.07$^{+0.13}_{-0.18}$    \smallskip \\ 
\ion{O}{VIII}                 & 16.03$^{+0.16}_{-0.17}$  & 15.71$^{+0.18}_{-0.23}$   &  15.92$^{+0.18}_{-0.19}$ & 15.70$^{+0.14}_{-0.17}$    \smallskip \\ 
\ion{O}{V}                    & 15.17$^{+0.26}_{-0.45}$  &  --                    &  15.28$^{+0.28}_{-0.45}$  & $\le$~14.40     \smallskip \\ 
\ion{O}{IV}                   & 15.61$^{+0.15}_{-0.21}$  &  --                    &  15.68$^{+0.16}_{-0.22}$  & $\le$~15.38     \smallskip \\ 
{\it \ion{O}{IV} (CIE)}       & 15.56$^{+0.21}_{-0.29}$   &  --                     & 15.51$^{+0.23}_{-0.35}$           & --              \smallskip \\
\ion{O}{I}                    & --                    &  --                     & 16.54$^{+0.11}_{-0.35}$   & 16.19$^{+0.10}_{-0.11}$   \smallskip \\ 
\ion{N}{VI}                   & 15.05$^{+0.25}_{-0.38}$  &  14.84$^{+0.25}_{-0.46}$  & 15.23$^{+0.33}_{-0.41}$   & $\le$~14.87     \smallskip \\ 
\ion{N}{I}                    & --                     &  --                    & 15.81$^{+0.12}_{-0.15}$   & $\le$~15.56      \smallskip \\ 
\ion{C}{VI}                   & 15.34$^{+0.19}_{-0.24}$  &  15.41$^{+0.13}_{-0.16}$  & 14.97$^{+0.25}_{-0.51}$   & $\le$~15.22     \smallskip \\ 
                              &                        &                        &               &                        \\ 
\hline\hline                  
\end{tabular}
\tablefoot{
\tablefoottext{a}{The values are obtained by fitting the data with the ``slab'' model, i.e. without an assumption about the ionisation balance, except for 
\ion{O}{IV}, for which we show also the value obtained under CIE assumption. 
The uncertainties include the systematic uncertainty of 1~m\AA\ and the allowance of v$_{tot}$~=~20--35~km~s$^{-1}$.}\\
\tablefoottext{b}{The value for RGS1, RGS2 and HRC is the weighted mean of the Ly$\alpha$ and Ly$\beta$ measurements. RGS2 only covers the  Ly$\beta$ line.}
}
}
\end{table}

\begin{table*}
 \centering
  \caption{COS column densities and velocities for the absorbing components in the \pks\ sight line \citep[][for high ions]{2012ApJ...749..157W}, Wakker et al., in prep. for low ions and \ion{H}{I} column density from FUSE)
  \label{cos.tab}}
    \begin{tabular}{ccccccccc}
  \hline\hline
      &        &                 &                 &                 &                     &                &                &                \\ 
v (km s$^{-1}$) & \multicolumn{8}{c}{$\log{N(\mathrm{cm}^{-2})}$ } \\
      & \ion{H}{I}     &   \ion{C}{II}           &  \ion{C}{IV}            & \ion{O}{VI}             & \ion{Si}{II}          & \ion{Si}{III}   &  \ion{Si}{IV}         & \ion{Al}{II}           \\
      &                      &                   &                 &                 &                     &                &                &                \\ 
-270  & 15.0$\pm$0.3         &  $<$12.55         & 13.36$\pm$0.04  & 13.22$\pm$0.08  & $<$11.96       & 12.15$\pm$0.10 & 12.34$\pm$0.22 & $<$12.00       \\
-232  & 14.6$\pm$0.3         &  $<$12.55         & 13.27$\pm$0.05  & 13.26$\pm$0.08  & $<$11.96       & 11.93$\pm$0.13  & 12.37$\pm$0.22 & $<$11.98       \\
-134  & 16.25$\pm$0.25       & 13.72$\pm$0.03    & 13.26$\pm$0.05  & 13.82$\pm$0.03  & 12.92$\pm$0.06 & 13.00$\pm$0.04 & 12.63$\pm$0.11 & 12.05$\pm$0.13 \\
MWY\tablefootmark{a}& 20.11  & $>$14.94          & 13.70$\pm$0.04  & 14.04$\pm$0.04  & 14.15$\pm$0.02 & $>$13.69       & 13.07$\pm$0.02 & $>$13.33             \\
      &        &                 &               &                 &                 &                &                &                \\ 
\hline\hline
 \end{tabular}
\tablefoot{\\
\tablefoottext{a}{The low velocity component}\\
}
\end{table*}

\section{CIE v.s. photo-ionisation}
\subsection{Method}
\label{CLOUDY}
The two competing ionisation processes in a case of TTG are photoionisation and collisional ionisation. It is likely that collisional ionisation equilibrium (CIE) holds for the HH due to its relatively high temperature. The column densities of ions occurring in TTG may be affected by photoionisation if the ionisation parameter was sufficiently high. Before using the ion column densities for estimating the temperature of TTG we made an attempt to determine the importance of photoionisation.

For this, we used the CLOUDY \citep{1998PASP..110..761F} code to assess the circumstances under which photoionisation might be important for the creation of  \ion{O}{IV} and \ion{O}{V}. 
CLOUDY assumes pure photoionisation and predicts the ionic column densities as a function of distance, metallicity and the ionisation parameter $U$ = $n$($\gamma$)/$n$(H), i.e. the ratio of number densities of the ionising photons to hydrogen.
Assuming the values for the distance and the metallicity, matching the FUV-measured ion column densities (see below) with the CLOUDY model
yields the estimate for $U$.  
An additional estimate of $n$($\gamma$) (see below) then yields the estimate for the hydrogen number density n(H), and consequently for the number densities of 
metals. The measurement of hydrogen column density N(\ion{H}{I}) (see below) then yields the path length L = N(H) / n(H) of the absorber and thus calculates the predictions for the 
column densities of the elements not used above. In practice, we use the above procedure to compute the predictions for the column densities of 
\ion{O}{IV} and \ion{O}{V}.

We then compare the predictions with our X-ray measurements of \ion{O}{IV} and \ion{O}{V} column densities as derived from RGS1 data without an assumption on the ionisation equilibrium (see Table~\ref{EWtau.tab}). The predictions vary depending on the assumed distance D and thus we will test different scenarios for 
the TTG origin: Galactic Halo/disk (D = 1 kpc), vicinity of the Galactic Halo (D = 20 kpc) and Local Group environment (D = 200 kpc).

To characterize the ionising radiation field in these different scenarios (i.e. the $n$($\gamma$) value at the different distances) , 
we followed the method of \cite{2005ApJ...630..332F}. They used the
\cite{2001cghr.confE..64H} model of the intensity and spectrum of the extragalactic ionising
radiation, combined with a model of the Galactic contribution, with the latter
varying as a function of the location of the cloud in the Galactic halo.

\subsection{Velocity structure}
A problem in the assessment of the role of the photoionisation is that in the direction of \pks\ , FUSE
and STIS spectra show a complicated kinematic structure 
\citep{1999ApJ...515..108S, 2003ApJS..146....1W, 2012ApJ...749..157W}.
\cite{1999ApJ...515..108S} reported the \ion{Si}{II} and \ion{C}{IV} and a
non-detection for \ion{C}{II} in the \pks\ sight line. They concluded that the clouds have 
$\log{U} \sim$ -3.0, density $\log{n ({\rm cm}^{-3})} \sim$ -4 and a size of several kpc. They interpreted these components as
analogous to those seen in Damped Ly-alpha Absorbers that originate in the
diffuse outer halos of galaxies. However, they did not include the Galactic
contribution to the ionising radiation field. Doing so would imply $\log{n}~\sim$-2 and a cloud size of a few 100 pc.

\cite{1999ApJ...515..108S} also derived a metallicity of $\log{Z}$\;=\;$-$0.5 for these clouds, but this is rather
uncertain. To make a prediction for ionic column densities we do our
calculations assuming possible metallicities $\log{Z}$\;=\;0,\;-0.5\;and\;-1.0. 

We improve significantly upon the previous results by using a more sensitive (as-yet unpublished) COS
spectrum, obtained in 2012, which reveals even more absorption components
than were analysed in the earlier papers. 
A full analysis of the COS spectra that shows these absorption lines lies
outside the scope of this paper. However, we present the column densities
measured for the different ions in Table~\ref{cos.tab}, since they are relevant for
understanding the \ion{O}{IV} and \ion{O}{V} column densities.

\begin{figure*}
\vspace{-8cm}
\hspace{-2cm}
\includegraphics[width=21cm,angle=0]{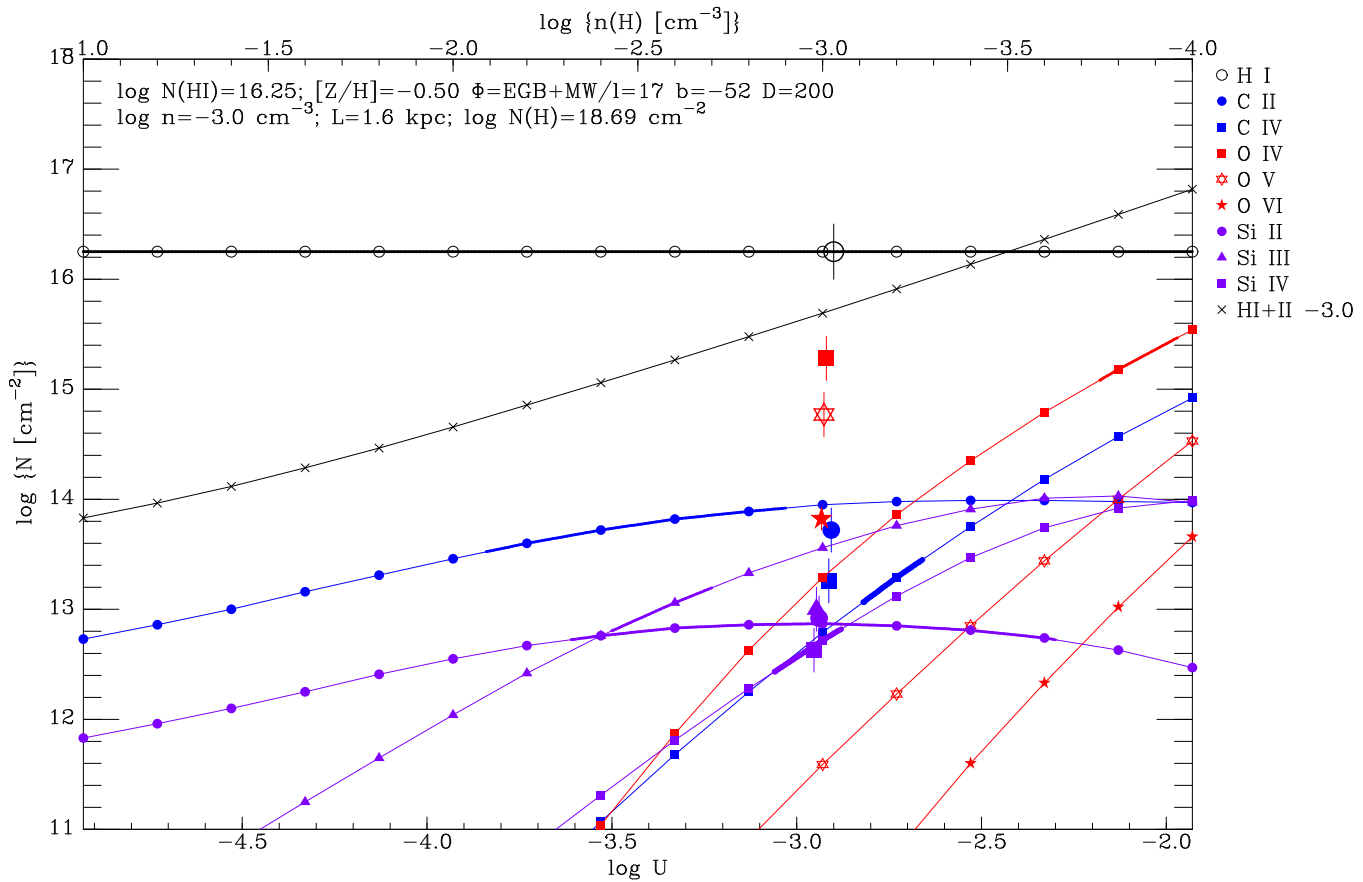}
\vspace{-8cm}
\caption{
Predicted ionic column densities as function of ionisation parameter U,
calculated for a location 200 kpc from the Sun in the direction of PKS2155-304.
Black lines/points are for hydrogen, blue for carbon, red for oxygen, purple
for silicon. Circles are for \ion{C}{II} and \ion{Si}{II}, triangles for \ion{Si}{III}, squares
for \ion{C}{IV} and \ion{O}{IV} and stars indicate \ion{O}{V}. The observed values are indicated
by the larger symbols near $\log{U}$ = -3.0, at which value both \ion{C}{IV} and \ion{Si}{IV}
match the observations. The thick parts of the lines show
where the prediction matches the range of observed values.}
\label{cloudy1.fig}
\end{figure*}

In the  \pks\ sight line, there are two components at velocities $<$75 km s$^{-1}$ and three at high velocity
(i.e. HVCs, the high-velocity clouds). The
low-velocity components are seen in low, intermediate and high ions. 
The low ions are likely associated with the nearby neutral disk and lower
halo which has a scale height of 3 kpc \citep{2009ApJ...702.1472S, 2012ApJ...749..157W}.
The high-ionisation lines (\ion{C}{IV}, \ion{O}{VI}) most likely have a mixed origin,
with \ion{O}{VI} probably originating in interfaces between hot and cool gas in
the lower and upper galactic halo. Some of the \ion{C}{IV} is also produced in these
regions, but another (unknown) fraction is due to photoionization. For the
modelling discussed below we assume the extreme situation in which all \ion{C}{IV}
is due to photoionization. A lower fraction would make the conclusions we derive stronger.

The two high-velocity components centered at $v$ = -270 and -232 km s$^{-1}$ are only
detected in lines of highly-ionised atoms (\ion{O}{VI}, \ion{C}{IV}, \ion{Si}{III} and \ion{Si}{IV}).  The
component at -134 km s$^{-1}$  shows these ions, as well as low-ionisation
lines of \ion{O}{I}, \ion{C}{II}, \ion{Si}{II}, and \ion{Al}{II}. 
The FUSE spectrum (Wakker et al., in prep.)  also shows Lyman series absorption for these clouds,
which allows the derivation of the \ion{H}{I} column densities shown in Table~\ref{cos.tab}.

Since the \ion{O}{IV} and \ion{O}{V} X-ray absorption is not kinematically resolved, we
cannot determine which fraction of the absorption is associated with the
low-velocity or the high-velocity gas. 
We will consider different possibilities below.

\subsection{HVC origin}
The location of the high-velocity clouds in the direction toward
\pks\ is uncertain. \cite{1999ApJ...515..108S} suggested they might
be rather distant (a few 100 kpc) clouds in the Local Group environment.
Alternatively, they might be in the vicinity of the Galactic halo at a distance of tens
of kpc. Considering their velocity and likely density it seems unlikely
that they are in the lower halo.

We first examined the lower distance scenario in which the -134 km s$^{-1}$ cloud is near the Milky Way, at a distance of 20 kpc, using the \ion{C}{II}, \ion{C}{IV}, \ion{Si}{II}, \ion{Si}{III} and \ion{Si}{IV} column densities measured for this component (see Table~\ref{cos.tab}).
We adopted $\log{\ion{H}{I}}$ = 16.25 from the FUSE spectrum (Wakker et al., in prep.) and assumed a metallicity $\log{Z}$\;=\;-0.5. In this case the CLOUDY modelling yields an ionization parameter $\log{U} \sim$ -2.7, and implies a total hydrogen column 
$\log{N(H)(\mathrm{cm^{-2}})}$\;=\;18.94.
This can be turned into the hydrogen number density and path
length by using the intensity of the ionization radiation field. In
our model (identical to that of Fox et al. 2004), at this location
(20 kpc from the Sun in the direction l,b=52,-17), this yields
$\log{n(H)(\mathrm{cm^{-3}}) }$\;$\sim$\;-2.0 and a path length of 280 pc.

Assuming instead that the absorber is at the distance of 200\;kpc, i.e. at the Local Group environment, we obtained
$\log{U}$\;$\sim$\;$-$3.0,  $\log{N(H)(\mathrm{cm^{-2}})}$\;=\;18.69, $\log{n(H)({\rm cm}^{-3})}$\;$\sim$\;-3.0, and a path length $\sim$1.6 kpc
(see Fig.~\ref{cloudy1.fig}).

In both cases N(\ion{O}{IV}) and N(\ion{O}{V}) are underpredicted by a factor of 100-1000 (see Fig.~\ref{cloudy1.fig}). We note that below the \ion{He}{II} limit
(54.4 eV), the Galactic contribution to the intensity of the ionizating radiation is about a factor ten larger than the extragalactic contribution,
while between 54 and $\sim$100 eV it is a factor two larger. Thus, at larger distances the radiation field is harder and the relative amount of
\ion{O}{IV} and \ion{O}{V} (ionization potentials 54.9 and 77.4 eV) is larger. However, even including the strong X-ray intensity present in some models of
the EGB (see e.g. \citet{2014ApJ...796...49S} for a comparison) is insufficient to produce the amount of observed \ion{O}{IV}, \ion{O}{V} and \ion{O}{VI} through 
photoionization.

We then experimented by varying the metallicity between $\log{Z}$\;=\;0 and $-$1 and replacing the column densities with those measured for the -270 and -232 km s$^{-1}$ clouds. Consequently, the cloud densities varied in the range $\log{n ({\rm cm}^{-3})}$ = -3.2 to -1.8 and the cloud sizes ranged from
0.1 to 5.1\;kpc. Yet, the \ion{O}{IV} and \ion{O}{V} underprediction remained at the same level as above.

Alternately, explaining N(\ion{O}{IV}) and N(\ion{O}{V}) as due to pure photoionization would require a
much higher ionization parameter ($\log{U} >-2.0$; implying a cloud size of 30\;kpc or larger). However, given the observed value of N(\ion{H}{I}),
in this case the predicted values of N(\ion{C}{IV}) and N(\ion{Si}{IV}) are more than two orders of magnitude larger than the
observed values.

In summary, it seems likely that the observed \ion{C}{II}, \ion{C}{IV}, \ion{Si}{II}, \ion{Si}{III} and \ion{Si}{IV} 
originate in a sub-kpc size photoionized condensation in the
Galactic vicinity or Local Group environment, but it is not possible that the X-ray-observed \ion{O}{IV} and \ion{O}{V} 
column densities can be produced in such a cloud.

\begin{figure*}
\vspace{-8.0cm}
\hspace{-2.0cm}
\includegraphics[width=21cm,angle=0]{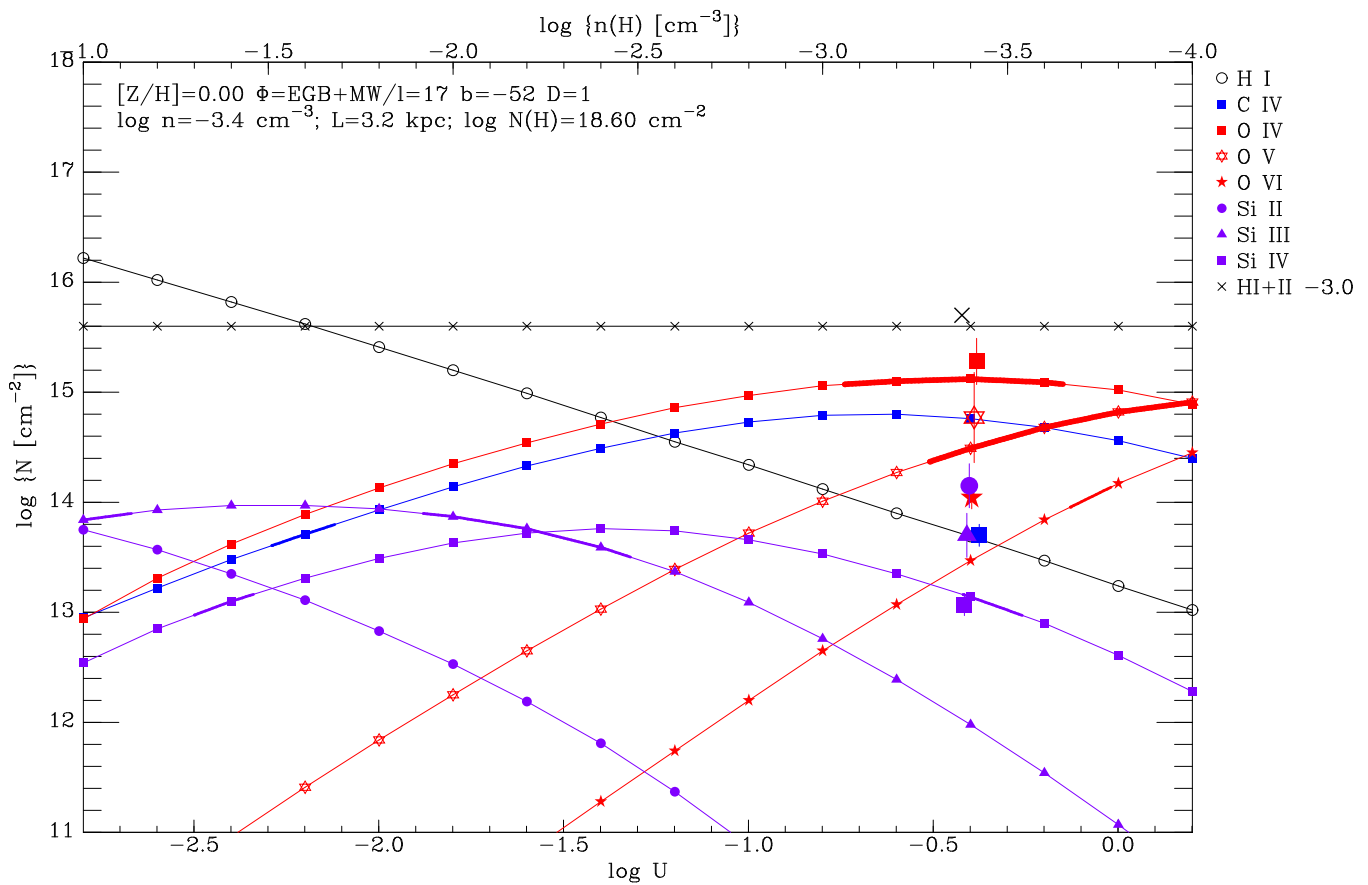}
\vspace{-8cm}
\caption{As Fig. \ref{cloudy1.fig} but for the Galactic Halo/disk case}
\label{cloudy2.fig}
\end{figure*}

\subsection{Galactic disk/halo origin}
Trying to assess whether photoionisation could be responsible for producing \ion{O}{IV}
and \ion{O}{V} in the Galactic disk or halo is more difficult, as it is unclear what the
corresponding column densities of kinematically resolved ions are. Most of the
\ion{H}{I} emission detected at 21~cm originates within a few 100~pc from the disk,
where the volume density is so high that \ion{O}{I} is the dominant oxygen ion while
\ion{O}{IV} and \ion{O}{V} will be mostly absent. It is also not possible to determine the fraction
of any other ion's column density that comes from the lower halo rather than the
disk. 

Our approach is to assume that \ion{O}{IV} and \ion{O}{V} originate in the same
volume and see whether the observed column densities can be consistent with
some value of N(H) and other ionic column densities for the ionising radiation
field near the disk. Since in ions such as \ion{C}{IV} and \ion{O}{VI} there are two components
with similar column densities \citep{2012ApJ...749..157W}, we use half the
observed value for the modelling. 

As shown in Fig.\ref{cloudy2.fig}, matching the observed
\ion{O}{IV}/\ion{O}{V} column densities and their ratio then is possible with an ionisation parameter
$\log{U} \sim$ -0.4
, and a total hydrogen density 
$\log{n({\rm H} ({\rm cm}^{-3}))} \sim$ -3.6 and a column density $\log{N({\rm H} ({\rm cm}^{-2}))} \sim$ 18.6. This implies
a path length of 3.2 kpc,
which would imply that the Galactic halo is filled with \ion{O}{IV}/\ion{O}{V} containing gas without much density fluctuations.
Such a situation can also fit the \ion{Si}{IV} and \ion{O}{VI} column densities,
but it underpredicts $N$(\ion{Si}{III}) by two orders of magnitude, which could come from a different phase.

However, this situation also predicts that $\log{N(\ion{C}{IV})}$ = 14.7, a factor ten larger
than is observed. Therefore, assuming that \ion{O}{IV}, \ion{O}{V} originate in photoionized
gas in the Galactic disk/halo environment is inconsistent with the data.

{\it Thus, we rule out a significant contribution by photoionisation to the \ion{O}{IV} and  \ion{O}{V}
column densities measured in our X-ray spectra.}

\begin{figure*}
\vspace{-8cm}
\hspace{-3cm}
\includegraphics[width=23cm,angle=0]{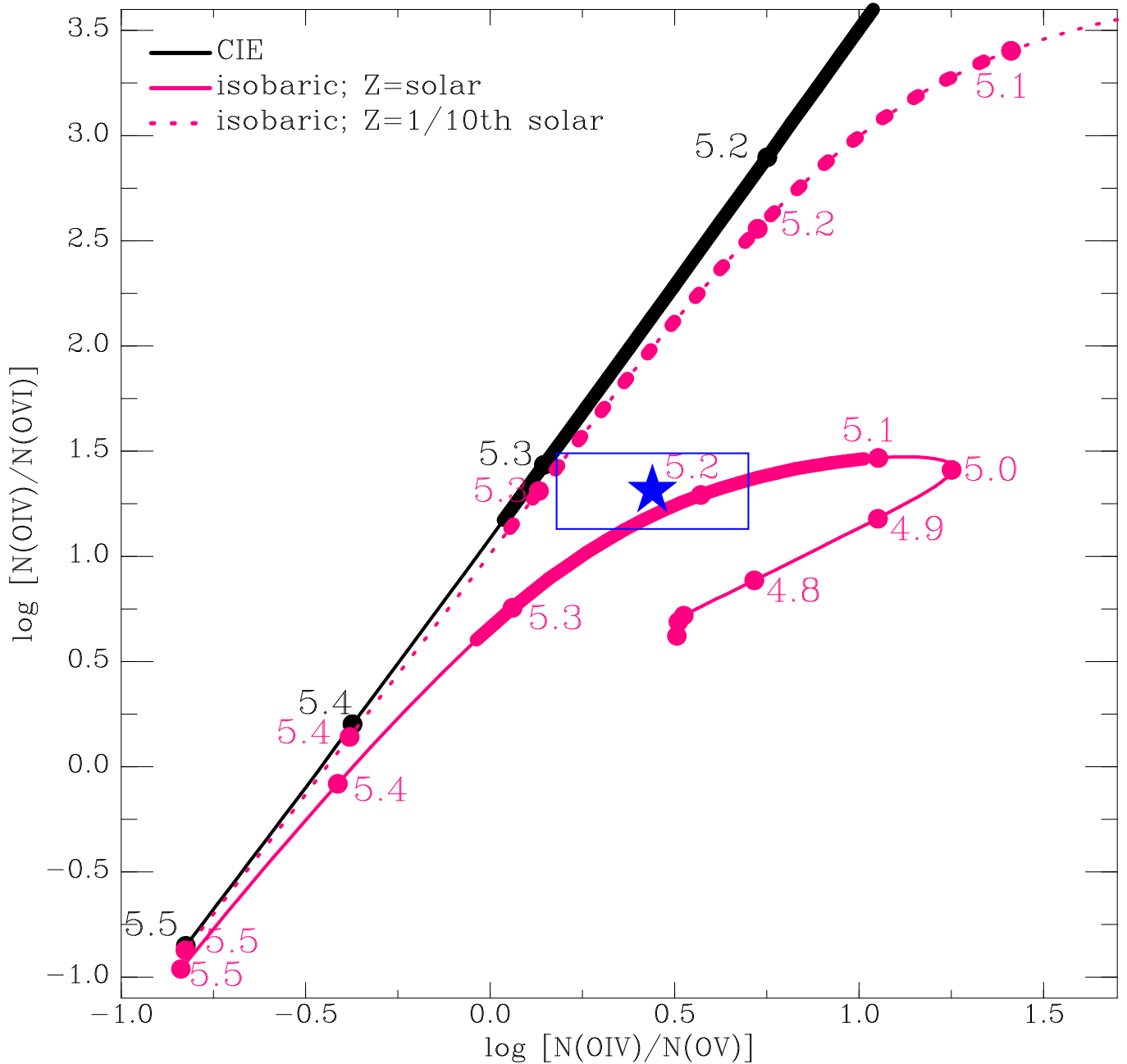}
\vspace{-4cm}
\caption{The column density ratio \ion{O}{IV}/\ion{O}{V} v.s. \ion{O}{IV}/\ion{O}{VI} based on the models of \cite{2007ApJS..168..213G}. The black line corresponds to CIE, while the red lines indicate the isobaric non-equilibrium cooling models using solar metallicity (solid line) and 0.1 $\times$ solar metallicity (dotted line). Corresponding $\log{T(\mathrm{K})}$ values are given by the labels along the curves. The curves are drawn as a thick line for the $\log{T(\mathrm{K})}$ values corresponding to the 1$\sigma$ 
confidence range obtained from the RGS1 SPEX CIE fit to the \ion{O}{IV}/\ion{O}{V} line ratio data, extended with the systematic offset of 0.05 dex between the different CIE implementations in \cite{2007ApJS..168..213G}  and SPEX. The blue star and box indicate the measurement.}
\label{instab.fig}
\end{figure*}

\section{Thermal instability}
\label{instab.sec}
Before reporting in detail the CIE modelling of the TTG, we study here the effects of the possible deviation of TTG from the ionisation equilibrium.
The CIE modelling indicated a temperature $\log{T(\mathrm{K})}$\;$\sim$\;5  
for TTG (see section~\ref{halo_model}), at which temperature the gas is in a thermally unstable phase, i.e. the heating and cooling time scales of TTG are shorter than the ionisation or recombination time scales  \citep[e.g.][]{2007ApJS..168..213G}.  Consequently, the TTG ions are associated with flows of heating or cooling material, as assumed in the Galactic Fountain models (see section~\ref{halo_model}). Thus, the CIE-derived temperature may be biased, given the suitable conditions.

Our X-ray measurements of \ion{O}{IV} and \ion{O}{V} column densities, derived from EW measurements without assumptions on the equilibrium,
cannot alone constrain the non-CIE models. Thus, we combined them with the FUV measurement of $\log{N(\ion{O}{VI})}$\;=\;14.31$\pm0.10$ in the  \pks\ sight line \citep{2012ApJ...749..157W}. In the case of \ion{O}{VI} about half the column density comes from the gas near $v$ $\sim$ 0 km~s$^{-1}$, while the other half is associated with high-velocity clouds in the line of sight. Although the \ion{O}{IV} and \ion{O}{V} X-ray absorption lines are unresolved, we assume a similar combination of low-velocity and high-velocity cooling parcels of transition temperature gas as in FUV.

We used the non-equilibrium cooling models of \cite{2007ApJS..168..213G} to calculate the column density ratios of \ion{O}{IV}, \ion{O}{V} and \ion{O}{VI} as a parcel of gas cools down from $T$\;=\;$10^6$\;to\;$10^{4.4}$\;K (see Fig.~\ref{instab.fig}). Between $\log{T(\mathrm{K})}$ = 6 and 5.3 the cooling is slow and the ionisation equilibrium is maintained. At $\log{T(\mathrm{K})}$\;$\sim$\;5.3 the cooling turns isobaric, i.e. the density of a TTG cloud increases as the temperature decreases. Consequently the \ion{O}{IV}/\ion{O}{VI} column density ratio falls below the CIE prediction for a given \ion{O}{IV}/\ion{O}{V} column density ratio. This reflects the "frozen-in" \ion{O}{VI} that cannot recombine fast enough as the temperature of the gas drops quickly. When the total column density $N$(H) exceeds 
\begin{equation}
N(\mathrm{H}) = 1.2 ~ \times ~ 10^{19} ~ \mathrm{cm}^{-2}~\times~\left[\frac{T}{10^5 ~\mathrm{K}}\right]^{2.0} 
\label{isochor.eq}
\end{equation}
the cooling becomes isochoric, i.e. the density stops growing and the pressure starts to decrease 
as the rate of cooling increases faster than the cloud can contract (see \cite{2007ApJS..168..213G} for a detailed description).
Using our CIE-derived\footnote{While it is in principle inconsistent to use $T$ and $N$(H) obtained under the CIE assumption to investige the boundary between the isobaric and isochoric phases, we show below that the deviation from the CIE is sufficiently small to produce a significant bias.} temperature estimate for the TTG (see section~\ref{halo_model}), the estimated $N$(H) is too small for the TTG to be in the isochoric phase.
Thus, in the following we only consider the CIE and isobaric phases.

The independent implementations of the CIE in \citet{2007ApJS..168..213G} and SPEX \citep{1996uxsa.conf..411K} may yield significant differences in the model 
predictions. In particular, in the SPEX package there is no option to use the solar abundance table of \citet{2009ARA&A..47..481A} which is implemented into the 
software based on \citet{2007ApJS..168..213G}. Instead, we use the abundance table of \citet{2009LanB...4B...44L} with SPEX.
 The cooling rate for a gas cloud with a given temperature and $N$(H) is different when the abundances of the important elements (O, C and Ne) are different. In fact we noticed that for a given $N$(\ion{O}{IV})/$N$(\ion{O}{V}) ratio in our measured range of [0.0,0.7], the CIE-derived $\log{T}$ of \citet{2007ApJS..168..213G} is about 
0.05 dex higher than that in SPEX (see Fig.~\ref{instab.fig}). This is comparable to the statistical uncertainties of our CIE 
temperature measurement. Thus, we consider this difference as an systematic uncertainty in the following.
 
The calculations indicated that the CIE model, as well as the isobaric model with abundance of 0.1 solar agreed only very marginally 
with the measured column density ratios of \ion{O}{IV}, \ion{O}{V} and \ion{O}{VI} (see Fig.~\ref{instab.fig}). 
On the other hand, the isobaric model with a solar abundance yields an excellent fit to the data with $\log{T({\mathrm K})}$ $\approx$5.2.
Our results support the Galactic Fountain model which predicts a solar abundance for the TTG and that the gas is in a cooling transition phase.

Yet the temperature is not low enough to produce a significant difference in the $N$(OIV)/$N$(OV) ratio for a given temperature in CIE and non-CIE models. Thus, there is no significant bias when deriving the TTG temperature via the X-ray spectral modelling assuming CIE (see next section).

\section{Local Hot Bubble}
\label{LHB}
Before modelling the Galactic absorption using separate components for the TTG and the hot halo, we checked whether the Local Hot Bubble (LHB) 
contributes significantly to our signal. The LHB surrounds the Sun with an extent of $\sim~$100~pc \citep{1977SSRv...20..815T, 1977ApJ...217L..87S, 1990ApJ...354..211S} and a spatially constant temperature of $kT$ = 0.10$\pm$0.01 keV (Liu et al. , 2016, in prep.).

Utilising the data from DXL mission \citep{2011ExA....32...83G}, Liu et al. (2016 in prep) removed the Solar Wind Charge Exchange (SWCX) emission 
from the RASS maps in the R12 band (0.1-0.3 keV). Since the LHB dominates the emission in this band (the hot Galactic halo being too hot)
these maps represent the clean LHB emission.
These maps indicate that in the PKS direction the LHB emission measure is
\begin{equation}
\mathrm{EM}_{\mathrm{LHB}}~ \equiv~n_e ~ \times ~ n_p ~ \times ~ l_{\mathrm{LHB}} ~ =  ~ 2.6 ~ \times ~ 10^{-3} ~ {\mathrm{cm}}^{-6} ~ {\mathrm{pc}},
\end{equation}
where $l_{\mathrm{LHB}}$ is the line-of-sight path length through LHB.

Since the sound crossing time in the LHB medium ($\sim$\thinspace$10^6$~years) is much smaller than the age of the LHB  ($\sim$\thinspace$10^7$ years), the LHB density is generally assumed to be spatially constant, measured as $n_e$\;=\;4.7$\pm0.5$\;$\times$\;$10^{-3}$\;cm$^{-3}$ by \cite{2014ApJ...791L..14S}.
Assuming the He to H ratio of 0.1 and complete ionisation for both H and He, i.e. that 
$n_e$\;=\;1.2\;$\times$\;$n_p$, 
we can thus estimate the total hydrogen column density of the LHB as
\begin{eqnarray}
N(\mathrm{H,LHB}) = n_e \times l_{\mathrm{LHB}} = 1.2 \times \frac{\mathrm{EM}_{\mathrm{LHB}}}{n_e} = \\
1.2 \times \frac{2.6\times 10^{-3} ~ {\mathrm{cm}}^{-6} ~ {\mathrm{pc}}}{4.7 \times 10^{-3} ~ {\mathrm{cm}}^{-3}} \approx 2.0 \times 10^{18} ~ {\mathrm{cm}}^{-2}.
\end{eqnarray}
This is less than 10\% of the equivalent hydrogen column density of the hot halo absorber we derived in this work (see below).
We estimated the effect of this component by additional CIE absorber, with a hydrogen column density fixed to the above value and the temperature to 0.1 keV.
There was a negligible change in the best-fit parameters of the hot halo and transition temperature gas. Thus, we do not include the LHB component in the following analysis.

\section{Galactic CIE absorption modelling}
\label{halo_model}
Since photoionisation can be excluded as the origin of the \ion{O}{IV} and \ion{O}{V} absorption (section~\ref{CLOUDY}), and the deviation from equilibrium does not introduce significant bias (section~\ref{instab.sec}), and the LHB does not contaminate the signal significantly, we proceed to assume that these ions reside in a gas in collisionally ionised equilibrium.

\subsection{Method}
\label{ciemeth}
In practice, we adopted the ``hot'' model of SPEX which calculates the transmission of a plasma with a given temperature ($kT$), equivalent hydrogen column density
($N$(H)), metal abundance (Z), metal-to-H ratio and non-thermal velocity ($v_{\mathrm{nt}}$), assuming CIE, as a function of the wavelength.
Multiplied with a given background emission spectrum, the model thus produces a prediction for the absorption lines. 
We found that the absorption due to the intervening Local Hot Bubble in this direction is negligible compared to the HH component (see section~\ref{LHB}).
We thus employed two ``hot'' components, one corresponding to the hot upper Galactic halo (HH) and one to the transition temperature gas (TTG) to fit the full 9--36 \AA\ band spectra. 

We adopted the element number density ratios from \cite{2009LanB...4B...44L}. 
We used the above discussed range of 15--35\;km\;s$^{-1}$ (see section \ref{velocity}) as a prior to the non-thermal velocities of each absorbing component, independently from each other.

For the LETG, we additionally included the neutral Galactic disc component (ND) which has been used to correct for the cold Galactic absorption in the RGS data (see Section~\ref{RGSdata}). We attempted the modelling of the blazar emission continuum in the LETG data with variants of the power-law model, but did not find a well-fitting solution. We found that a spline model with eight terms described the full band continuum acceptably. 

For a given individual line, the equivalent hydrogen column density and the metal abundance are fully anti-correlated. 
Our data were not adequate to break this degeneracy and thus we further assumed in our fits that the metal abundances are solar.

In summary, our essential choices for the analysis were:\\
1) The collisional ionisation equilibrium is valid (as indicated by the CLOUDY analysis). \\
2) The abundances of the metals in HH and TTG are solar.\\
3) The non-thermal velocity dispersion $\sigma_V$ is in the range 15--35\;km\;s$^{-1}$ (as indicated by the FUV analysis in section \ref{velocity} ).

Thus, our free parameters in the fits are the temperature $kT$ and the equivalent hydrogen column density $N$(H) of the two components, and the 
continuum normalisation. For LETG, the shape of the continuum is also allowed to vary.
The comparison of the above parameters obtained using RGS2 with those obtained with the other instruments is complicated. Namely, RGS2 does not cover the wavelengths of the OIV and OV lines and thus cannot constrain the TTG component. Also, RGS2 does not cover the \ion{O}{VII} 1s-2p line. As a result, the derived parameters had unusefully large uncertainties and differences between those derived with RGS1 and HRC. Thus, we do not discuss the RGS2 measurements further for the CIE analysis. Given the problems of the systematically inconsistent line measurements with LETG/ACIS-S (see section~\ref{ACIS_instr}), in particular the non-detection of the \ion{O}{IV} and \ion{O}{V} lines of the TTG component, we also excluded the LETG/ACIS-S data from the further analysis.

Considering a single line with a temperature near the value of the peak strength, higher column densities can to an extent compensate for lower ionisation factors when reproducing the given line data. Thus, we expect some level of correlation between the temperature and the equivalent hydrogen column density in the spectral analysis.  In order to consider this correlation (and possible correlations between the different absorbing components) we adopted the following procedure to 
estimate the best-fit parameters and their uncertainties for each absorber (HH and TTG), and for the consequent \ion{O}{IV} and \ion{O}{V} column densities for 
TTG. The procedure was applied independently to the RGS1 and HRC data. 

\begin{enumerate}

\item
We obtained the global best-fit $N$(H)-$kT$ combination by fitting the full band data with the two-temperature CIE absorber model (plus a fixed neutral absorber for HRC) using a $\chi^2$ minimisation (thus yielding $\chi^2_{\mathrm{min, global}}$ to be used below).\\

\item
We then determined dense 2D grids around the best-fit values in the $N$(H)\;-\;$kT$ parameter space for both HH and TTG.\\

\item
We fixed the $N$(H)\;-\;$kT$ value pair of a given component to one defined by the above grid.\\

\item
We re-fitted the data with $N$(H)\;-\;$kT$ of the other component, as well as the normalisation of the continuum as free parameters, thus obtaining
$\chi^2_{\mathrm{min,i}}$ . \\

\item
We recorded the difference $\Delta \chi^2$ between the $\chi^2$ values of the above best fit ($\chi^2_{\mathrm{min,i}}$) and the global best-fit ($\chi^2_{min,global}$).\\

\item
We repeated steps 2-5, but this time choosing a different combination of $N$(H)\;-\;$kT$ value pair.\\ 

\item
We repeated steps 2-6 but this time for the other absorber.\\

\item
When all the value combinations in the pre-defined grid had been used, we constructed 2D $\Delta \chi^2$ maps corresponding to our $N$(H)\;-\;$kT$ grids for HH and TTG.\\

\item
We determined the 1$\sigma$ sub-space of the grid by selecting only such value combinations which yielded $\Delta \chi^2 \le 2.3$.\\

\item
We used the maximal variation of the projection of the allowed parameter sub-space into $N$(H) and $kT$ axes as 1$\sigma$ uncertainty intervals.\\

\item
Using each best fit with the $N$(H)\;-\;$kT$ value combinations within the allowed  1$\sigma$ sub-space for TTG we computed the distribution of the allowed \ion{O}{IV} and \ion{O}{V} column densities and adopted the extremal values as the 1$\sigma$ CL interval. 

\end{enumerate}

\subsection{Results}
\label{bestfit}
\begin{figure*}
\vspace{-15cm}
\includegraphics[width=22cm,angle=0]{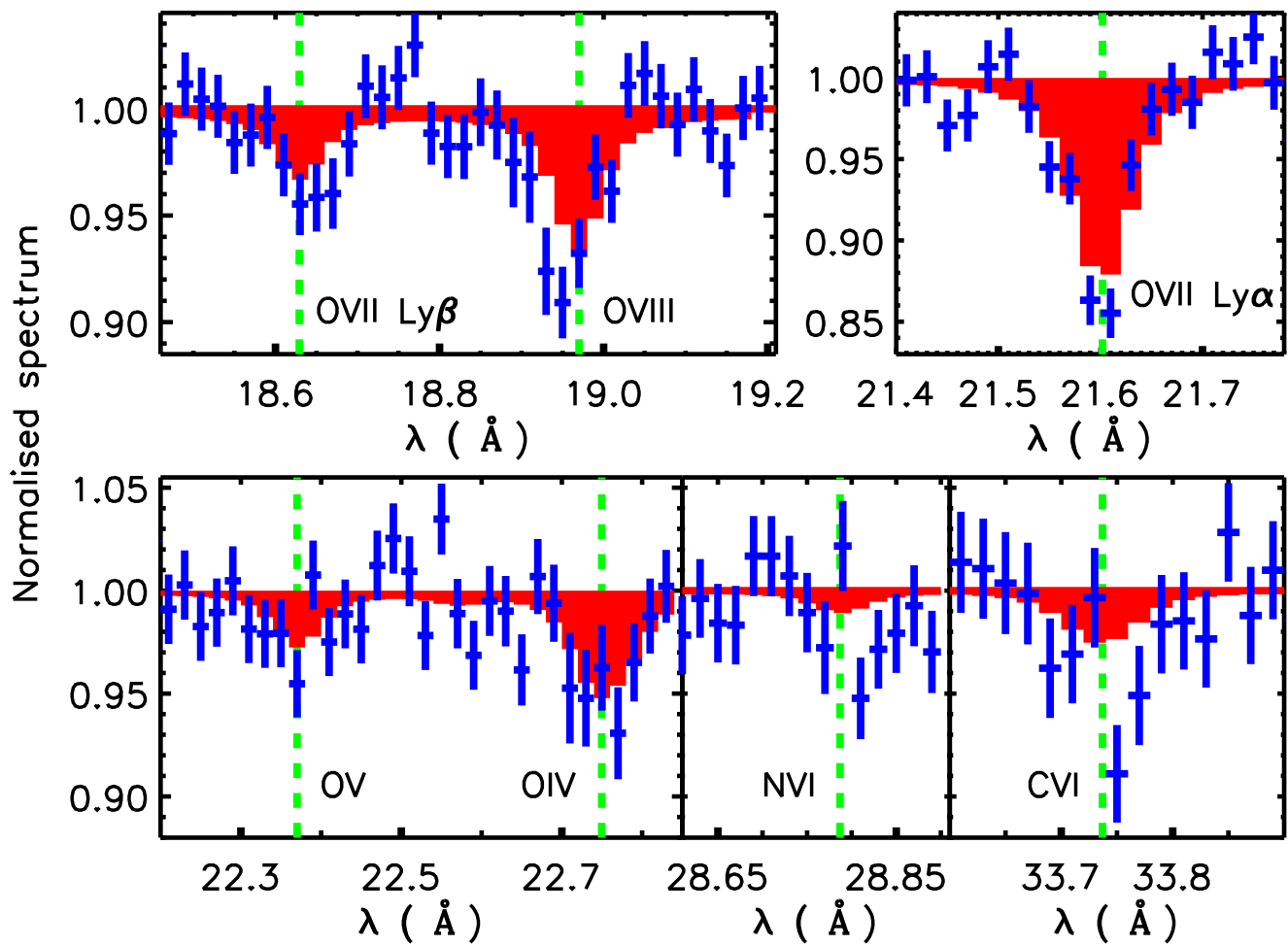}
\caption{The best-fit CIE model (red lines) around the lines significantly detected with RGS1 data (blue crosses).}
\label{rsg1_cie.fig}
\end{figure*}

The best-fit multi-component CIE models (see Table~\ref{phys.tab}) described the data reasonably well (see  Fig.~\ref{rsg1_cie.fig} for the RGS1 fits). 
However, the \ion{C}{VI} and \ion{N}{VI} lines at the longest wavelengths were not well modelled. This may indicate the existence 
of an additional component of the hot halo with a slightly lower temperature than reported above. We will investigate this issue further in a future work.

\subsubsection{TTG}
\label{TTG}
\begin{figure*}
\vspace{-15cm}
\includegraphics[width=22cm,angle=0]{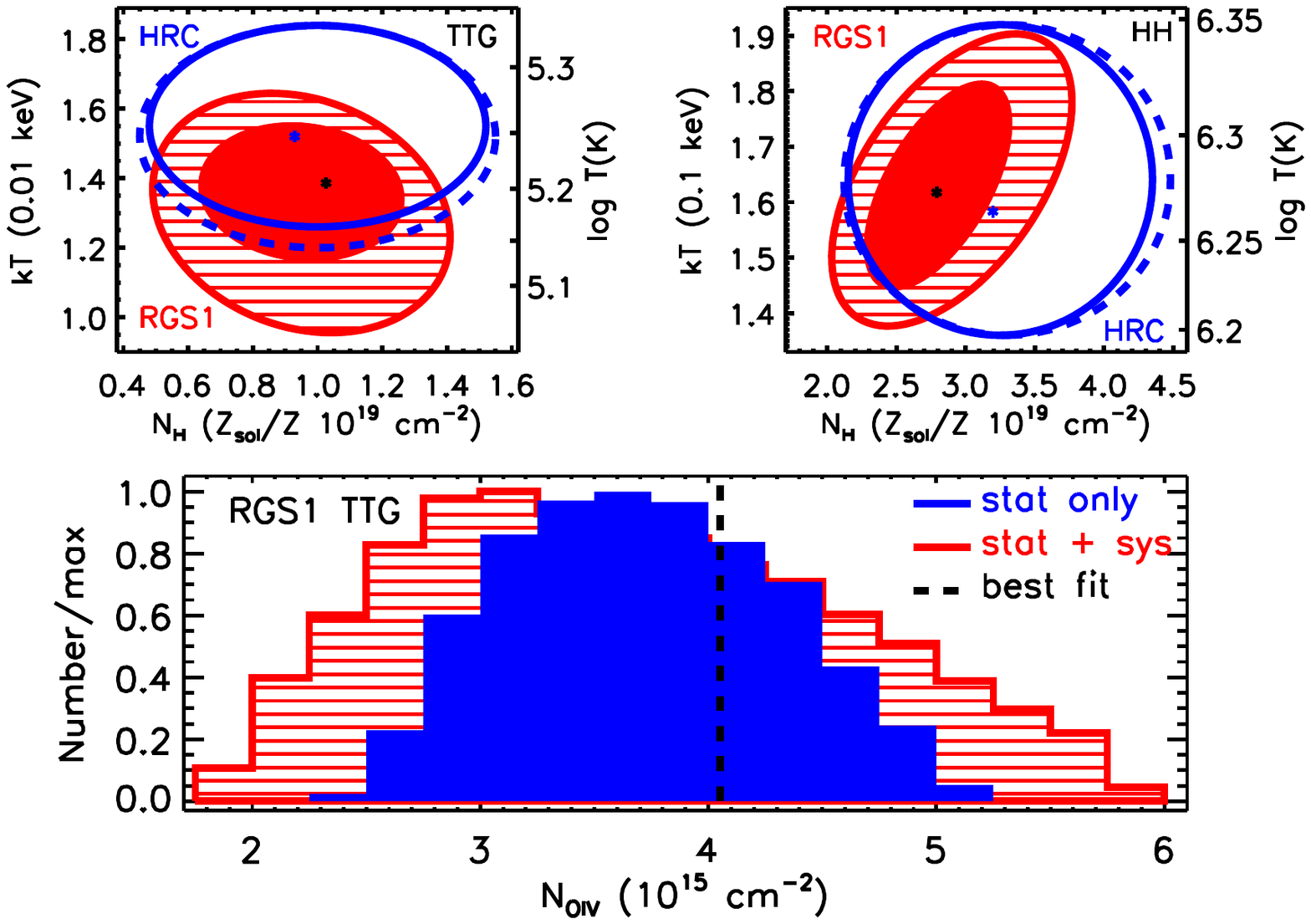}
\caption{The 1$\sigma$ uncertainties in the kT-N$_H$ plane for TTG (upper left panel) and HH (upper right panel) for RGS1 (red) and HRC(blue).
The results obtained when using only statistical uncertainties are indicated with the filled ellipses (RGS1) and solid blue lines (HRC).
The red shaded regions and the blue dashed line indicate the results obtained when including the 2\% uncertainty for the effective area calibration uncertainty.
The global best-fit spectral model parameters are indicated as black (RGS1) and blue (HRC) asterisks.
The lower panel shows the distribution of the \ion{O}{IV} column density of TTG, corresponding to the allowed RGS1 kT-N$_H$ region in the upper left panel.
The results obtained with only statistical uncertainties are indicated with a blue region while the red region shows the results when including the 2\% systematics. The dashed line indicates the value corresponding to the global best-fit spectral model. 
}
\label{T_NH.fig}
\end{figure*}

Using only the statistical uncertainties the RGS1 data constrained the TTG parameters as  
$N$(H$_{\mathrm{TTG}}$) = 1.0[0.6$-$1.3] $\times$ $10^{19}$ $\frac{Z_{\odot}}{Z_{\mathrm{TTG}}}$ cm$^{-2}$ and 
$kT_{\mathrm{TTG}}$ = 1.4[1.2$-$1.5] $\times$ $10^{-2}$\ keV 
($\log{T_{\mathrm{TTG}}({\mathrm{K}})}$\;=\;5.2[5.1$-$5.3]). These are in excellent agreement with the independent HRC measurements:
$N$(H$_{\mathrm{TTG}}$) = 0.9[0.5$-$1.5] $\times$ $10^{19}$ $\frac{Z_{\odot}}{Z_{\mathrm{TTG}}}$ cm$^{-2}$ and 
$kT_{\mathrm{TTG}}$ = 1.5[1.3$-$1.8] $\times$ $10^{-2}$ keV 
($\log{T_{\mathrm{TTG}}({\mathrm{K}})}$ = 5.2[5.1$-$5.3], 
see Fig.~\ref{T_NH.fig}). 

At the above temperature \ion{O}{IV} is the dominating oxygen ion.
The \ion{O}{IV} column density distributions are smooth, but in case of RGS1 it peaks at a slighly smaller value than the one corresponding to the global best-fit model (see section~\ref{ciemeth}).
We thus adopted the peak location as the most likely value for RGS1, obtaining consistent values 
$N$(OIV,RGS1) = 3.6[2.4$-$5.1] $\times$ $10^{15}$ cm$^{-2}$ (see Fig.~\ref{T_NH.fig}) and 
$N$(OIV,HRC)  = 3.3[1.6-5.3] $\times$ $10^{15}$ cm$^{-2}$.

The above consistencies imply that the uncertainties of the calibration of the effective areas
of RGS1 and HRC (see section~\ref{RGSdata}) do not introduce a significant bias to our results of the CIE analysis of the TTG.  
However, we must estimate the effect of the calibration uncertainties to the error bars of the parameters derived above.
Since we do not know the probability distributions of the calibration uncertainties of both the RGS1 and the HRC, we cannot perform a proper error propagation analysis.
Rather, we approximated the effect by adding 2\% of the best-fit model flux to the statistical uncertainties 
at each spectral bin, and repeating the above analysis (see section~\ref{ciemeth}).
This had a significant effect on RGS1: the 1$\sigma$ uncertainty intervals increased from
[0.6$-$1.3] to [0.5$-$1.4] $\times$ $10^{19}$ $\frac{Z_{\odot}}{Z_{\mathrm{TTG}}}$ cm$^{-2}$
and 
[1.2$-$1.5] to [0.9$-$1.6] $\times$ $10^{-2}$ keV 
for 
$N$(H$_{\mathrm{TTG}}$) and $kT_{\mathrm{TTG}}$, respectively.

The inclusion of the systematic uncertainties rendered the RGS1 \ion{O}{V} column density rather uncertain.
However, we obtained useful constraints for the RGS1 \ion{O}{IV} column density:
$N$(OIV,RGS1) = 3.6[1.9$-$5.8] $\times$ $10^{15}$ cm$^{-2}$ (see Fig.~\ref{T_NH.fig}), consistent with the HRC value of 
$N$(OIV,HRC)  = 3.3[1.5$-$5.5] $\times$ $10^{15}$ cm$^{-2}$.
{\it The values are consistent with those derived without an assumption on the ionisation balance (see section~\ref{NOCIE} and Table~\ref{EWtau.tab})}. In addition to the results of the CLOUDY analysis (see section~\ref{CLOUDY}) this indicates that the photo-ionisation is not significantly contaminating the TTG absorption signal. 

The larger statistical uncertainties of the HRC data render the calibration uncertainty effect negligible on the results derived with 
the HRC. Due to the larger statistical HRC uncertainties and the significant level of systematic uncertainties in the RGS1 data,
the weighted averages of RGS1 and HRC values are less precise than the results obtained with RGS1 alone (using 2\% systematic uncertainties
and symmetrising the confidence interval):

\begin{eqnarray}
N(\mathrm{H}_{\mathrm{TTG}}) & = &  1.0\pm0.5~\times~10^{19}~\frac{Z_{\odot}}{Z_{\mathrm{TTG}}}~{\mathrm{cm}}^{-2} \\
kT_{\mathrm{TTG}}           &  = & 1.4\pm0.3~\times~10^{-2}~{\mathrm{keV}} \\
\log{T_{\mathrm{TTG}}({\mathrm{K}})} & = & 5.2\pm0.1 \\
N(\ion{O}{IV}) & = &  3.6\pm2.0~\times~10^{15}~{\mathrm{cm}}^{-2}
\end{eqnarray}

\subsubsection{HH}
We repeated the analysis of the previous section, but this time for the Hot Halo (HH) component, including the 2\% systematic uncertainties.
The results for RGS1 and HRC are consistent (see Table~\ref{phys.tab}). As in the case of TTG, the weighted averages of RGS1 and HRC measurements are less precise than 
those obtained with RGS1 alone, which yielded:
 
\begin{eqnarray}
N(H_{\mathrm{HH}}) & = &  2.7\pm0.8~\times~10^{19}~\frac{Z_{\odot}}{Z_{\mathrm{TTG}}}~{\mathrm{cm}}^{-2} \\
kT_{\mathrm{HH}}        &  = &   1.6\pm0.2~\times~10^{-1}~{\mathrm{keV}} \\
\log{T_{\mathrm{HH}}(\mathrm{K})} & = & 6.3\pm0.1 \\
\end{eqnarray}

\begin{table}
 \centering
  \caption{The results of multi-temperature CIE modelling\tablefootmark{a}
  \label{phys.tab}}
\scriptsize{
    \begin{tabular}{lcccc}
  \hline\hline
  &                       &                &         \\ 
  &  $N_\mathrm{H}$       & $N_{\ion{O}{IV}}$ & $kT$ & $\log{T ({\rm K})}$        \\
  &  ($10^{19}~\frac{Z_{\odot}}{Z_{TTG}}$ cm$^{-2}$) &  ($10^{15}$ cm$^{-2}$)   & (keV)  &      \\               
  &                       &                &    &      \\ 
  & \multicolumn{4}{c}{RGS1} \\
     &                    &     &             &             \\ 
TTG  &   1.0$\pm$0.5       &  3.6$\pm$2.0 &   1.4$\pm$0.3~$\times~10^{-2}$  & 5.2$\pm$0.1 \\
HH   &   2.7$\pm$0.8       & --  &   1.6$\pm$0.2~$\times~10^{-1}$  & 6.3$\pm$0.1 \\
     &               & &                               &               \\ 
 & \multicolumn{4}{c}{HRC} \\
     &             &   &                                &                                \\ 
TTG  & 0.9$\pm$0.5 &  3.3$\pm$2.0 & 1.5$\pm$~0.3$\times~10^{-2}$ &  5.2$\pm$0.1   \\ 
HH   & 3.1$\pm$1.2 & ---& 1.6$\pm$0.3~$\times~10^{-1}$   &   6.3$\pm$0.1  \\ 
     &               & &                               &      \\ 
         \hline\hline
 \end{tabular}
\tablefoot{\\
\tablefoottext{a}{Including systematic 2\% uncertainties}\\
}}
\end{table}

\section{Conclusions}
We analysed $\sim$~3~Ms of data of \pks\ obtained with high resolution X-ray spectrometers RGS1 and RGS2 on-board XMM-Newton and LETG/HRC-S and LETG/ACIS-S on-board Chandra. The main conclusions are listed below.\\

\begin{itemize}

\item
Due to the high statistical quality of the data, and due to the coverage of the important wavelengths with more than one instrument, we obtained secure detection of ten Galactic absorption lines at CL higher than 99.75\% in the sight line towards \pks\ . Due to our very strict criteria for line detection, we may have omitted some important lines at problematic wavelengths. \\

\item
We discovered significant absorption blend from inner transitions of \ion{O}{IV}: 
1s-2p $^2$S (22.571~\AA ), 1s-2p $^2$P (22.741~\AA ) and  1s-2p $^2$D (22.777~\AA ), and from
\ion{O}{V}  1s-2p (22.370~\AA ), consistently with RGS1 and LETG/HRC-S.  \\ 

\item
Combining our X-ray measurements of \pks\ with those obtained in FUV we determined that the \ion{O}{IV} and \ion{O}{V} absorption cannot be produced by photoionisation, since the observed column densities would imply an ionisation parameter that would overpredict the FUV-observed \ion{C}{IV} column density by an order of magnitude.  \\ 

\item
Using non-CIE cooling models of \cite{2007ApJS..168..213G} we showed that the column density ratios of \ion{O}{IV}, \ion{O}{V} and \ion{O}{VI}
are in excellent agreement with a model where all these ions originate in isobarically cooling gas with solar abundances and a temperature of $\log{T(K)}$ $\sim$ 5.2. 
The CIE model and an isobaric 0.1 solar model are strongly disfavored by the data.
We thus identify this absorber with the transition temperature gas TTG as predicted by the Galactic fountain theory,
previously detected only in FUV \citep[e.g.][]{2012ApJ...749..157W}. \\

\item
The temperature derived from the non-CIE modelling is not low enough to produce a significant difference between 
the temperature based on the \ion{O}{IV}/\ion{O}{V} column density ratio assuming CIE or non-CIE.  \\ 

\item
Assuming CIE, the spectral fit to the \pks\ spectrum, essentially to the \ion{O}{IV}/\ion{O}{V} ratio data, 
yielded a temperature of
$kT$~=~1.4$\pm$0.3~$\times~10^{-2}$~keV , i.e. 
$\log{T(\mathrm{K})}~=~5.2\pm0.1$ for TTG. \\

\item
The column density of \ion{O}{IV}, obtained with CIE assumption ($N_{\mathrm{OIV}}$~=~3.6[1.9--5.8]~$\times~10^{15}$~cm$^{-2}$) agrees with that derived 
directly from the EW measurement of the line, without assumptions on the phase of the gas. \\

\item 
The equivalent hydrogen column density corresponding to the \ion{O}{IV} absorber is  1.0$\pm0.5~\times~10^{19}~\frac{Z_{\odot}}{Z_{TTG}}$ cm$^{-2}$.  \\

\item
The significant discrepancy in the measured EW of the residual signal at the instrumental feature at $\lambda$~=~23.35~\AA\  
indicates that the signal cannot be purely due to astrophysical \ion{O}{II}, as suggested by \cite{2016MNRAS.457..676N}.

\end{itemize}

\begin{acknowledgements}
JN is funded by PUT246 grant from Estonian Research Council. Support for BPW was provided by NASA through grants HST-GO-13892.01-A, 
HST-GO-13840.05-A and HST-GO-13721.02-A from the Space Telescope Science Institute, which is operated by the Association of Universities for 
Research in Astronomy. We thank J. de Plaa, J. Drake and E. Saar for help.

\end{acknowledgements}

\bibliographystyle{aa} 
\bibliography{halobib} 

\begin{appendix}

\section{Observation log}
We provide here information about the XMM-Newton and Chandra observations used in this work.

\begin{table*}
 \centering
  \caption{RGS observations
  \label{rgs.tab}}
    \begin{tabular}{ccccc}
  \hline\hline
            &                       &         & &               \\ 
Orbit & obsid      &  norm\tablefootmark{a} & $\Gamma$\tablefootmark{b} & $t$\tablefootmark{c} (ks)  \\
            &                       &         & &               \\ 
   174 & 0080940101 &   4069 &  2.69 &   59\\
  174 & 0080940301 &   3288 &  2.76 &   60\\
  362 & 0124930301 &   8558 &  2.65 &   90\\
  450 & 0124930501 &   3782 &  2.55 &  102\\
  545 & 0124930601 &   2867 &  2.59 &  114\\
  724 & 0158960101 &   2120 &  2.85 &   27\\
  908 & 0158960901 &   2397 &  2.87 &   29\\
  908 & 0158961001 &   3217 &  2.77 &   40\\
  993 & 0158961101 &   4442 &  2.64 &   29\\
 1095 & 0158961301 &   4714 &  2.71 &   60\\
 1171 & 0158961401 &   1836 &  2.69 &   65\\
 1266 & 0411780101 &   2682 &  2.64 &  100\\
 1349 & 0411780201 &   4613 &  2.77 &   68\\
 1543 & 0411780301 &   5463 &  2.68 &   61\\
 1734 & 0411780401 &   3578 &  2.87 &   65\\
 1902 & 0411780501 &   1843 &  2.79 &   71\\
 2084 & 0411780601 &   3053 &  2.56 &   64\\
 2268 & 0411780701 &    708 &  2.85 &   57\\
 2449 & 0411782101 &   1579 &  2.74 &   76\\
 2542 & 0727770101 &   1724 &  2.60 &   95\\
 2631 & 0727770501 &   1900 &  2.76 &   88\\
 2633 & 0727770901 &   1676 &  2.82 &   64\\
 2632 & 0727771001 &   1498 &  2.74 &   39\\
 2632 & 0727771101 &   1879 &  2.83 &   39\\
 2726 & 0727771301 &   1336 &  2.64 &   92\\
      &            &        &       &        \\ 
\multicolumn{4}{c}{\bf Total} &       {\bf 1700 ks} \\
\hline\hline                  
\end{tabular}
\tablefoot{\\
\tablefoottext{a}{The normalisation of the absorbed power-law fitted to the RGS data over the full band; the units are 
$4\pi$ $\times$ the number of photons m$^{-2}$s$^{-1}$keV$^{-1}$ at 1 keV} \\
\tablefoottext{b}{The best-fit photon index of the power-law described above} \\
\tablefoottext{c}{The exposure time} \\
}
\end{table*}

\begin{table*}
 \centering
  \caption{LETG observations
  \label{letg.tab}}
    \begin{tabular}{ccc}
  \hline\hline
                &                       &                       \\ 
       OBS. ID.	&	Detector	&	Exposure (ks)	\\
                &                       &                       \\ 
	1015	&	ACIS-456789	&	9.5	\\
	10662	&	ACIS-456789	&	28.4	\\
	11965	&	ACIS-456789	&	28.4	\\
	13096	&	ACIS-456789	&	26.5	\\
	14265	&	ACIS-456789	&	26.5	\\
	15475	&	ACIS-456789	&	28.4	\\
	16423	&	ACIS-56789	&	28.1	\\
	1703	&	ACIS-456789	&	25.3	\\
	1790	&	ACIS-456789	&	20.9	\\
	1791	&	ACIS-456789	&	20.9	\\
	1792	&	ACIS-456789	&	20.9	\\
	1793	&	ACIS-456789	&	20.9	\\
	1794	&	ACIS-456789	&	20.9	\\
	1795	&	ACIS-456789	&	19.7	\\
	1796	&	ACIS-456789	&	19.5	\\
	1797	&	ACIS-456789	&	19.5	\\
	1798	&	ACIS-456789	&	19.5	\\
	1799	&	ACIS-456789	&	19.9	\\
	2323	&	ACIS-456789	&	9.0	\\
	2324	&	ACIS-456789	&	8.6	\\
	2335	&	ACIS-456789	&	29.1	\\
	3168	&	ACIS-456789	&	28.8	\\
	3667	&	ACIS-456789	&	14.2	\\
	3668	&	ACIS-456789	&	13.5	\\
	3669	&	ACIS-456789	&	42.4	\\
	3707	&	ACIS-456789	&	26.9	\\
	4416	&	ACIS-456789	&	46.5	\\
	6090	&	ACIS-456789	&	27.5	\\
	6091	&	ACIS-456789	&	29.2	\\
	6874	&	ACIS-456789	&	28.5	\\
	6924	&	ACIS-456789	&	9.5	\\
	6927	&	ACIS-456789	&	27.0	\\
	7293	&	ACIS-456789	&	8.5	\\
	8388	&	ACIS-456789	&	29.3	\\
	9704	&	ACIS-456789	&	27.8	\\
	9706	&	ACIS-456789	&	9.5	\\
	9708	&	ACIS-456789	&	9.2	\\
	9710	&	ACIS-456789	&	9.2	\\
	9713	&	ACIS-456789	&	29.7	\\
                &                       &               \\            
		&	{\bf Total}	& {\bf 900}	\\
                &                       &               \\                     
                &                       &               \\      
	1013	&	HRC-S	&	26.6	\\
	3709	&	HRC-S	&	13.7	\\
	7294	&	HRC-S	&	10.0	\\
	1704	&	HRC-S	&	25.9	\\
	7295	&	HRC-S	&	10.1	\\
	4406	&	HRC-S	&	13.8	\\
	8379	&	HRC-S	&	29.8	\\
	5172	&	HRC-S	&	26.9	\\
	9707	&	HRC-S	&	10.1	\\
	3166	&	HRC-S	&	29.8	\\
	6922	&	HRC-S	&	10.0	\\
	9709	&	HRC-S	&	10.1	\\
	331	&	HRC-S	&	62.7	\\
	6923	&	HRC-S	&	29.9	\\
                &               &               \\ 
		& {\bf Total}	&  {\bf 300}	\\
                &               &               \\ 
\hline\hline                  
\end{tabular}
\end{table*}

\clearpage

\end{appendix}

\end{document}